\documentclass[sigconf]{acmart}
\usepackage{blindtext}
\usepackage[utf8]{inputenc}
\usepackage[T1]{fontenc}
\usepackage[draft,inline,nomargin,index]{fixme}
\usepackage{array,amssymb,amsmath,graphicx,comment,multirow,microtype}
\usepackage{booktabs,color,xspace,balance,csquotes}
\usepackage{mathtools}
\usepackage{colortbl}

\newcommand{\ra}[1]{\renewcommand{\arraystretch}{#1}}
\renewcommand\footnotetextcopyrightpermission[1]{}

\fxsetup{theme=colorsig,mode=multiuser,inlineface=\itshape,envface=\itshape}
\FXRegisterAuthor{ps}{aps}{Piotr}
\FXRegisterAuthor{am}{aam}{Alan}

\definecolor{applegreen}{rgb}{0.55, 0.71, 0.0}

\newcommand{\para}[1]{{\vspace{7pt} \bf \noindent #1 \hspace{10pt}}}

\newenvironment{packed_itemize}{
\begin{itemize}
  \setlength{\itemsep}{2pt}
  \setlength{\parskip}{0pt}
  \setlength{\parsep}{0pt}
  \setlength{\topsep}{2pt}
  \setlength{\itemindent}{0pt}
}{\end{itemize}}

\newenvironment{packed_enumerate}{
\begin{enumerate}
  \setlength{\itemsep}{2pt}
  \setlength{\parskip}{0pt}
  \setlength{\parsep}{0pt}
  \setlength{\topsep}{2pt}
  \setlength{\itemindent}{0pt}
}{\end{enumerate}}

\newcommand{\eg}{e.g.,\ }
\newcommand{\etal}{et~al.\xspace}
\newcommand{\ie}{i.e.,\ }

\newcommand{\aka}{a.k.a.\ }

\newcommand{\fulltitle}{Discrimination through optimization: \\How Facebook's ad delivery can lead to skewed outcomes}
\newcommand{\fullkeywords}{information retrieval; fairness;}

\definecolor{linkColor}{RGB}{6,125,233}
\definecolor{left}{RGB}{100,149,237}
\definecolor{right}{RGB}{205,92,92}
\hypersetup{%
  pdftitle={\fulltitle},
  pdfauthor={},
  pdfkeywords={\fullkeywords},
  bookmarksnumbered,
  pdfstartview={FitH},
  colorlinks,
  citecolor=black,
  filecolor=black,
  linkcolor=black,
  urlcolor=linkColor,
  breaklinks=true,
  hypertexnames=false
}

\begin{document}

\fancyhead{}

\renewcommand{\sectionautorefname}{\S}
\renewcommand{\subsectionautorefname}{\S}
\renewcommand{\subsubsectionautorefname}{\S}

\title{\fulltitle}

\author{Muhammad Ali$^*$}
\affiliation{\institution{Northeastern University}}
\email{mali@ccs.neu.edu}

\author{Piotr Sapiezynski$^*$}
\affiliation{\institution{Northeastern University}}
\email{sapiezynski@gmail.com}

\author{Miranda Bogen}
\affiliation{\institution{Upturn}}
\email{miranda@upturn.org}

\author{Aleksandra Korolova}
\affiliation{\institution{University of Southern California}}
\email{korolova@usc.edu}

\author{Alan Mislove}
\affiliation{\institution{Northeastern University}}
\email{amislove@ccs.neu.edu}

\author{Aaron Rieke}
\affiliation{\institution{Upturn}}
\email{aaron@upturn.org}

\begin{abstract}
The enormous financial success of online advertising platforms is partially due to the precise targeting features they offer. Although researchers and journalists have found many ways that advertisers can target---or exclude---particular groups of users seeing their ads, comparatively little attention has been paid to the implications of the platform's {\em ad delivery} process, comprised of the platform's choices about which users see which ads.

It has been hypothesized that this process can ``skew'' ad delivery in ways that the advertisers do not intend, making some users less likely than others to see particular ads based on their demographic characteristics. In this paper, we demonstrate that such skewed delivery occurs on Facebook, due to market and financial optimization effects as well as the platform's own predictions about the ``relevance'' of ads to different groups of users. We find that both the advertiser's budget and the content of the ad each significantly contribute to the skew of Facebook's ad delivery.
Critically, we observe significant skew in delivery along gender and racial lines for ``real'' ads for employment and housing opportunities despite neutral targeting parameters.

Our results demonstrate previously unknown mechanisms that can lead to potentially discriminatory ad delivery, even when advertisers set their targeting parameters to be highly inclusive. This underscores the need for policymakers and platforms to carefully consider the role of the ad delivery optimization run by ad platforms themselves---and not just the targeting choices of advertisers---in preventing discrimination in digital advertising.\footnote{The delivery statistics of ad campaigns described in this work can be accessed at https://facebook-targeting.ccs.neu.edu/ \\$^*$ These two authors contributed equally}

\end{abstract}

\maketitle
\pagenumbering{arabic}
\section{Introduction}

Powerful digital advertising platforms fund most popular online services today, serving ads to billions of users daily.
At a high level, the functionality of these advertising platforms can be divided into two phases: {\em ad creation}, where advertisers submit the text and images that comprise the content of their ad and choose targeting parameters, and {\em ad delivery}, where the platform delivers ads to specific users based on a number of factors, including advertisers' budgets, their ads' performance, and the predicted relevance of their ads to users.

One of the underlying reasons for the popularity of these services with advertisers is the rich suite of {\em targeting} features they offer during ad creation, which allow advertisers to precisely specify which users (called the {\em audience}) are eligible to see the advertiser's ad.
The particular features that advertisers can use for targeting vary across platforms, but often include demographic attributes, behavioral information, users' personally identifiable information (PII), mobile device IDs, and web tracking pixels~\cite{andreou-2018-explanations,venkatadri-2018-targeting}.

Due to the wide variety of targeting features---as well as the availability of sensitive targeting features such as user demographics and interests---researchers have raised concerns about discrimination in advertising, where groups of users may be excluded from receiving certain ads based on advertisers' targeting choices~\cite{speicher-2018-targeted}.
This concern is particularly acute in the areas of credit, housing, and employment, where there are legal protections in the U.S. that prohibit discrimination against certain protected classes in advertising~\cite{FairHousingActDiscrimination,AgeDiscriminationinEmploymentAct,EqualCreditOpportunityAct}.
As ProPublica demonstrated in 2016~\cite{FacebookExcludeRace}, this risk is not merely theoretical: ProPublica investigators were able to run housing ads that explicitly excluded users with specific ``ethnic affinities'' from receiving them.\footnote{In response, Facebook banned the use of certain attributes for housing ads, but many other, un-banned, mechanisms exist for advertisers that achieve the same outcome~\cite{speicher-2018-targeted}.  Facebook agreed as part of a lawsuit settlement stemming from these issues to go further by banning age, gender, and certain kinds of location targeting---as well as some related attributes---for housing, employment, or credit ads~\cite{FacebookHousingCreditEmployment}.}  
Recently, the U.S. Department of Housing and Urban Development (HUD) sued Facebook over these concerns and others, accusing Facebook's advertising platform of ``encouraging, enabling, and causing'' violations of the Fair Housing Act~\cite{FacebookHUDLawsuit}. 

\para{The role of ad delivery in discrimination}
Although researchers and investigative journalists have devoted considerable effort to understanding the potential discriminatory outcomes of ad targeting, comparatively little effort has focused on ad delivery, due to the difficulty of studying its impacts without internal access to ad platforms' data and mechanisms.
However, there are several potential reasons why the ad delivery algorithms used by a platform may open the door to discrimination.  

\textit{First}, consider that most platforms claim their aim is to show users ``relevant'' ads: for example, Facebook states ``we try to show people the ads that are most pertinent to them''~\cite{FacebookRelevantAds}.
Intuitively, the goal is to show ads that particular users are likely to engage with, even in cases where the advertiser does not know a priori which users are most receptive to their message.
To accomplish this, the platforms build extensive user interest profiles and track ad performance to understand how different users interact with different ads. 
This historical data is then used to steer future ads towards those users who are most likely to be interested in them, and to users like them.
However, in doing so, the platforms may inadvertently cause ads to deliver primarily to a skewed subgroup of the advertiser's selected audience, an outcome that the advertiser may not have intended or be aware of.
As noted above, this is particularly concerning in the case of credit, housing, and employment, where such skewed delivery might violate antidiscrimination laws.

\textit{Second}, market effects and financial optimization can play a role in ad delivery, where different desirability of user populations and unequal availability of users may lead to skewed ad delivery~\cite{dwork-2018-fairness}.
For example, it is well-known that certain users on advertising platforms are more valuable to advertisers than others~\cite{ExpensiveKeywords,liu-2014-ads,saeztrumper-2014-value}.
Thus, advertisers who choose low budgets when placing their ads may be more likely to lose auctions for such ``valuable'' users than advertisers who choose higher budgets.
However, if these ``valuable'' user demographics are strongly correlated with protected classes, it could lead to discriminatory ad delivery due to the advertiser's budget alone.
Even though a low budget advertiser may not have intended to exclude such users, the ad delivery system may do just that because of the higher demand for that subgroup. 

Prior to this work, although hypothesized \cite{dwork-2018-fairness, lambrecht-2018-algorithmic, UpturnFacebookAmicusBrief}, it was not known whether the above factors resulted in skewed ad delivery in real-world advertising platforms. 
In fact, in response to the HUD lawsuit~ \cite{FacebookHUDLawsuit} mentioned above, Facebook claimed that the agency had ``no evidence'' of their ad delivery systems' role in creating discrimination~\cite{FacebookHUDLawsuitProPublica}.

\para{Contributions}
In this paper, we aim to understand whether ads could end up being shown in a skewed manner---i.e., where some users are less likely than others to see ads based on their demographic characteristics---due to the ad delivery phase alone. 
In other words, we determine whether the ad delivery could cause skewed delivery {\em that an advertiser did not cause by their targeting choices and may not even be aware of}.
We focus on Facebook---as it is the most mature platform offering advanced targeting features---and run dozens of ad campaigns, hundreds of ads with millions of impressions, spending over \$8,500 as part of our study.

Answering this question---especially without internal access to the ad delivery algorithm, user data, and advertiser targeting data or delivery statistics---involves overcoming a number of challenges. These include separating market effects from optimization effects, distinguishing ad delivery adjustments based on the ad's performance measured through user feedback from initial ad classification, and developing techniques to determine the racial breakdown of the delivery audience (which Facebook does not provide). 
The difficulty of solving these without the ad platform's cooperation in a rigorous manner may at least partially explain the lack of knowledge about the potential discriminatory effects due to ad delivery to date.
After addressing these challenges, we find the following:

\textit{First}, we find that {\em skewed delivery can occur due to market effects alone}.
Recall the hypothesis above concerning what may happen if advertisers in general value users differently across protected classes.
Indeed, we find this is the case on Facebook: when we run identical ads targeting the same audience but with varying budgets, the resulting audience of users who end up seeing our ad can range from over 55\% men (for ads with very low budgets) to under 45\% men (for ads with high budgets).

\textit{Second}, we find that {\em skewed delivery can occur due to the content of the ad itself} (\ie the ad headline, text, and image, collectively called the {\em ad creative}).
For example, ads targeting the same audience but that include a creative that would stereotypically be of the most interest to men (\eg bodybuilding) can deliver to over 80\% men, and those that include a creative that would stereotypically be of the most interest to women (\eg cosmetics) can deliver to over 90\% women. 
Similarly, ads referring to cultural content stereotypically of most interest to Black users (\eg hip-hop) can deliver to over 85\% Black users, and those referring to content stereotypically of interest to white users (\eg country music) can deliver to over 80\% white users, even when targeted identically by the advertiser.
Thus, despite placing the same bid on the same audience, the advertiser's ad delivery can be heavily skewed based on the ad creative alone.

\textit{Third}, we find that {\em the ad image itself has a significant impact on ad delivery}.
By running experiments where we swap different ad headlines, text, and images, we demonstrate that the differences in ad delivery can be significantly affected by the image alone. 
For example, an ad whose headline and text would stereotypically be of the most interest to men with the image that would stereotypically be of the most interest to women delivers primarily to women at the same rate as when all three ad creative components are stereotypically of the most interest to women. 

\textit{Fourth}, we find that {\em the ad image is likely automatically classified by Facebook}, and that this classification can skew delivery from the beginning of the ad's run.
We create a series of ads where we add an alpha channel to stereotypically male and female images with over 98\% transparency; the result is an image with all of the image data present, but that looks like a blank white square to humans.
We find that there are statistically significant differences in how these ads are delivered depending on the image used, despite the ads being visually indistinguishable to a human.
This indicates that the image classification---and, therefore, relevance determination---is likely an automated process, and that the skew in ad delivery can be due in large part to skew in Facebook's automated estimate of relevance, rather than ad viewers' interactions with the ad.

\textit{Fifth}, we show that {\em real-world employment and housing ads can experience significantly skewed delivery}.
We create and run ads for employment and housing opportunities, and use our methodology to measure their delivery to users of different races and genders.
When optimizing for clicks, we find that ads with the same targeting options can deliver to vastly different racial and gender audiences depending on the ad creative alone.
In the most extreme cases, our ads for jobs in the lumber industry reach an audience that is 72\% white and 90\% male, our ads for cashier positions in supermarkets reach an 85\% female audience, and our ads for positions in taxi companies reach a 75\% Black audience, even though the targeted audience specified by us as an advertiser is identical for all three.
We run a similar suite of ads for housing opportunities, and find skew there as well: despite the same targeting and budget, some of our ads deliver to an audience of over 72\% Black users, while others delivery to over 51\% Black users.
While our results only speak to how our particular ads are delivered (\ie we cannot say how housing or employment ads {\em in general} are delivered), the significant skew we observe even on a small set of ads suggests that real-world housing and employment ads are likely to experience the same fate.

Taken together, our results paint a distressing picture of heretofore unmeasured and unaddressed skew that can occur in online advertising systems, which have significant implications for discrimination in targeted advertising.
Specifically, due to platforms' optimization in the ad delivery stage together with market effects, 
ads can unexpectedly be delivered to skewed subsets of the advertiser's specified audience.
For certain types of ads, such skewed delivery might implicate legal protections against discriminatory advertising. 
For example, Section 230 of the U.S. Communications Decency Act (CDA) protects publishers (including online platforms) from being held responsible for third-party content.
Our results show Facebook's integral role in shaping the delivery mechanism and might make it more difficult for online platforms to present themselves as neutral publishers in the future.
We leave a full exploration of these implications to the legal community.
However, our results indicate that regulators, lawmakers, and the platforms themselves need to think carefully when balancing the optimization of ad platforms against desired societal outcomes, and remember that ensuring that individual advertisers do not discriminate in their targeting is insufficient to achieve non-discrimination goals sought by regulators and the public.

\para{Ethics} 
All of our experiments were conducted with careful consideration of ethics.
We obtained Institutional Review Board review of our study at Northeastern University (application \#18-11-13), with our protocol being marked as ``Exempt''.
We minimized harm to Facebook users when we were running our ads by always running ``real'' ads (in the sense that if people clicked on our ads, they were brought to real-world sites relevant to the topic of the ad).
While running our ads, we never intentionally chose to target ads in a discriminatory manner (\eg we never used discriminatory targeting parameters).
To further minimize the potential for discrimination, we ran most of our experimental ads in categories with no legal salience (such as entertainment and lifestyle); we only ran ad campaigns on jobs and housing to verify whether the effects we observed persist in these domains.
We minimized harm to the Facebook advertising platform by paying for ads and using the ad reporting tools in the same manner as any other advertiser.
The particular sites we advertised were unaffiliated with the study, and our ads were not defamatory, discriminatory, or suggestive of discrimination.

\smallskip

\section{Background}\label{sec:background}
Before introducing our methodology and analyses, we provide background on online display advertising, describe Facebook's advertising platform's features, and detail related work.

\subsection{Online display advertising}
Online display advertising is now an ecosystem with aggregate yearly revenues close to \$100 billion~\cite{OnlineAdvertisingBillions}.
The web advertising ecosystem is a complex set of interactions between ad publishers, ad networks, and ad exchanges, with an ever-growing set of entities involved at each step allowing advertisers to reach much of the web.
In contrast, online services such as Facebook and Twitter run advertising platforms that primarily serve a single site (namely, Facebook and Twitter themselves).
In this paper, we focus on single-site advertising platforms, but our results may also be applicable to more general display advertising on the web; we leave a full investigation of the extent to which this is the case to future work. 

The operation of platforms such as Facebook and Twitter can be divided into two phases: {\em ad creation} and {\em ad delivery}.
We provide more details on each below.

\para{Ad creation}
Ad creation refers to the process by which the advertiser submits their ad to the advertising platform.
At a high level, the advertiser has to select three things when doing so:
\begin{packed_enumerate}
\item{\em Ad contents}: Advertisers will typically provide the ad headline, text, and any images/videos. Together, these are called the {\em ad creative}.  They will also provide the link where the platform should send users who click.
\item{\em Audience Selection/Targeting}: Advertisers need to select which platform users they would like to see the ad (called the {\em audience}).
\item{\em Bidding strategy}: Advertisers need to specify how much they are willing to pay to have their ads shown.  This can come in the form of a per-impression or per-click bid, or the advertiser can simply place an overall {\em bid cap} and allow the platform to bid on their behalf.
\end{packed_enumerate}
Once the advertiser has entered all of the above information, they submit the ad for review;\footnote{Most platforms have a review process to prevent abuse or violations of their platforms' advertising policies~\cite{FacebookAdReview,TwitterAdReview}.} once it is approved, the ad will move to the ad delivery phase.

\para{Ad delivery}
Ad delivery refers to the process by which the advertising platform shows ads to users.
For every opportunity to show a user an ad (e.g., an \textit{ad slot} is available as the user is browsing the service), the ad platform will run an {\em ad auction} to determine, from among all of the ads that include the current user in the audience, which ad should be shown.

In practice, however, the ad delivery process is somewhat more complicated.
{\em First}, the platforms try to avoid showing ads from the same advertiser repeatedly in quick succession to the same user; thus, the platforms will sometimes disregard bids for recent winners of the same user.
{\em Second}, the platforms often wish to show users relevant ads; thus, rather than relying solely on the bid to determine the winner of the auction, the platform may incorporate a relevance score into consideration, occasionally allowing ads with lower bids but more relevance to win over those with higher bids.
{\em Third}, the platforms may wish to evenly spread the advertiser budget over their specified time period, rather than use it all at once, which introduces additional complexities as to which ads should be considered for particular auctions.
The exact mechanisms by which these issues are addressed are not well-described or documented by the platforms.

Once ads enter the ad delivery phase, the advertising platforms give advertisers information on how their ads are performing.
Such information may include detailed breakdowns (e.g., along demographic or geographic lines) of the characteristics of users to whom their ad is being shown and those who click on the ad.

\begin{figure}[t!]
	\centering
	\includegraphics[width=1\linewidth]{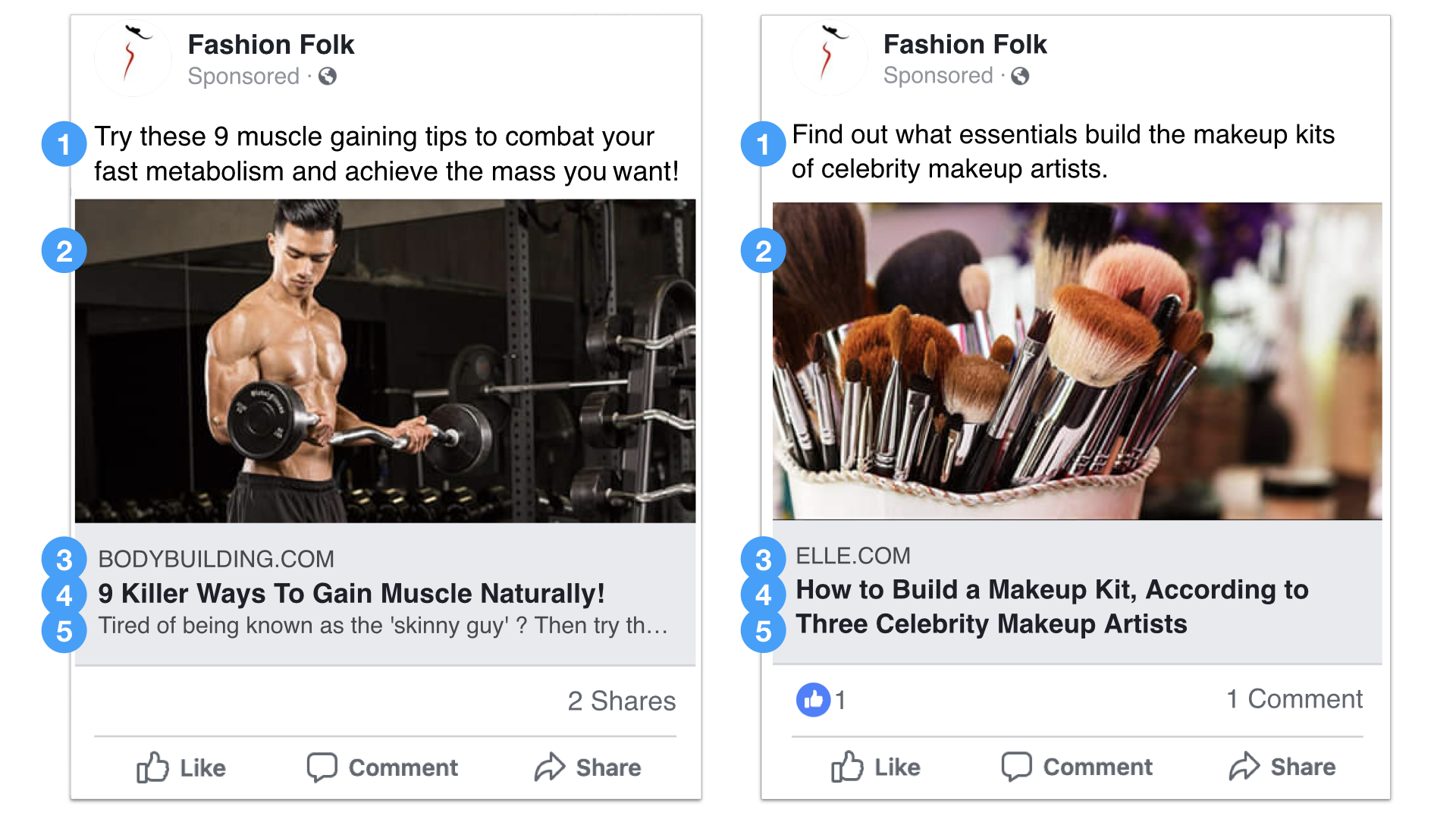}
	\caption{Each ad has five elements that the advertiser can control: (1) the ad text, entered manually by the advertiser, (2) the images and/or videos, (3) the domain, pulled automatically from the HTML \texttt{meta} property \texttt{og:site\_name} of the destination URL, (4) the title, pulled automatically from the HTML \texttt{meta} property \texttt{og:title} of the destination URL, and (5) the description from \texttt{meta} property \texttt{og:description} of the destination URL.  The title and description can be manually customized by the advertiser if they wish.}
	\label{fig:screenshot-bodybuilding}
\end{figure}

\subsection{Facebook's advertising platform}\label{subsec:fb-impl}

In this paper, we focus on Facebook's advertising platform as it is one of the most powerful and feature-rich advertising platforms in use today.
As such, we provide a bit more background here about the specific features and options that Facebook provides to advertisers.

\para{Ad contents}
Each ad placed on Facebook must be linked to a {\em Page}; advertisers are allowed to have multiple Pages and run ads for any of them.
Ads can come in multiple forms, such as promoting particular posts on the page.
However, for typical ads, the advertiser must provide (a) the headline and text to accompany the ad, and (b) one or more images or videos to show to the user.
Optionally, the advertiser can provide a {\em traffic destination} to send the user to if they click (e.g., a Facebook Page or an external URL); if the advertiser provides a traffic destination, the ad will include a brief description (auto-generated from the HTML \texttt{meta} data) about this destination.
Examples showing all of these elements are presented in Figure~\ref{fig:screenshot-bodybuilding}.

\para{Audience selection}
Facebook provides a wide variety of audience selection (or {\em targeting}) options~\cite{andreou-2019-facebook,ghosh-2019-facebook,speicher-2018-targeted,andreou-2018-explanations}.
In general, these options fall into a small number of classes:
\begin{packed_itemize}

\item{\em Demographics and attributes}: Similar to other advertising platforms~\cite{TwitterAdAttributes,GoogleAdAttributes}, Facebook allows advertisers to select audiences based on  demographic information (\eg age, gender, and location), as well as profile information, activity on the site, and data from third-parties.
Recent work has shown that Facebook offers over 1,000 well-defined attributes and hundreds of thousands of free-form attributes~\cite{speicher-2018-targeted}.

\item{\em Personal information}: Alternatively, Facebook allows advertisers to specify {\em the exact users} who they wish to target by either (a) uploading the users' personally identifiable information including names, addresses, and dates of birth~\cite{FacebookCustomAudiencesAPI,venkatadri-2019-pii,venkatadri-2018-targeting}, or (b) deploying web tracking pixels on third-party sites~\cite{FacebookPixelAudience}.
On Facebook, audiences created using either mechanism are called {\em Custom Audiences}.\footnote{Google, Twitter, and Pinterest all provide similar features; these are called {\em Customer Match}~\cite{GoogleCustomerMatch}, {\em Tailored Audiences}, and  {\em Customer Lists}~\cite{PinterestAudiences}, respectively.}

\item{\em Similar users}: Advertisers may wish to find ``similar'' users to those who they have previously selected.
To do so, Facebook allows advertisers to create {\em Lookalike Audiences}\footnote{Google and Pinterest offer similar features: on Google it is called {\em Similar Audiences}~\cite{GoogleSimilarAudience}, and on Pinterest it is called {\em Actalike Audiences}~\cite{PinterestActalikeAudience}.} by starting with a source Custom Audience they had previously uploaded; Facebook then ``identif[ies] the common qualities of the people in it'' and creates a new audience with other people who share those qualities~\cite{FacebookLookalikeAudience}.
\end{packed_itemize}
Advertisers can often combine many of these features together, for example, by uploading a list of users' personal information and then using attribute-based targeting to further narrow the audience.

\para{Objective and bidding}
Facebook provides advertisers with a number of {\em objectives} to choose from when placing an ad~\cite{FacebookAdObjectives}, where each tries to maximize a different {\em optimization event} the advertiser wishes to occur.
These include ``Awareness'' (simply optimizing for the most \textit{impressions}, \aka views), ``Consideration'' (optimizing for clicks, engagement, etc.), and ``Conversion'' (optimizing for sales generated by clicking the ad).
For each objective, the advertiser bids on the objective itself (e.g., for ``Awareness'', the advertiser would bid on ad impressions).
The bid can take multiple forms, and includes the start and end time of the ad campaign and either a lifetime or a daily budget cap.
With these budget caps, Facebook places bids in ad auctions on the advertisers' behalf.
Advertisers can optionally specify a per-bid cap as well, which will limit the amount Facebook would bid on their behalf for a single optimization event.

\para{Facebook's ad auction}
When Facebook has ad slots available, it runs an ad auction among the active advertisements bidding for that user.
However, the auction does not just use the bids placed by the advertisers; Facebook says~\cite{FacebookAdAuctions}:
\begin{displayquote}
The ad that wins an auction and gets shown is the one with the highest {\em total value} [emphasis added]. Total value isn't how much an advertiser is willing to pay us to show their ad. It's combination of 3 major factors: (1) Bid, (2) Estimated action rates, and (3) Ad quality and relevance.
\end{displayquote}
Facebook defines ``Estimated action rates'' as ``how well an ad performs'', meaning whether or not {\em users in general} are engaging with the ad~\cite{FacebookAdPrinciples}.
They define ``Ad quality and relevance'' as ``how interesting or useful we think a given user is going to find a given ad'', meaning how much {\em a particular user} is likely to be interested in the ad~\cite{FacebookAdPrinciples}.

Thus, it is clear that Facebook attempts to identify the users within an advertiser's selected audience who they believe would find the ad most useful (i.e., those who are most likely to result in an optimization event) and shows the ad preferentially to those users.
Facebook says exactly as such in their documentation~\cite{FacebookAdDelivery}:
\begin{displayquote}
During ad set creation, you chose a target audience ... and an optimization event ... We show your ad to people in that target audience who are likely to get you that optimization event
\end{displayquote}
Facebook provides advertisers with an overview of how well-matched it believes an ad is with the target audience using a metric called {\em relevance score}, which ranges between 1 and 10.  Facebook says~\cite{FacebookRelevantAds}:
\begin{displayquote}
Relevance score is calculated based on the positive and negative feedback we expect an ad to receive from its target audience.
\end{displayquote}
Facebook goes on to say~\cite{FacebookRelevantAds}:
\begin{displayquote}
Put simply, the higher an ad's relevance score, the less it will cost to be delivered.  This is because our ad delivery system is designed to show the right content to the right people, and a high relevance score is seen by the system as a positive signal.
\end{displayquote}

\para{Statistics and reporting}
Facebook provides advertisers with a feature-rich interface~\cite{FacebookAdsManager} as well as a dedicated API~\cite{FacebookMarketingAPI} for both launching ads and monitoring those ads as they are in ad delivery.
Both the interface and the API give semi-live updates on delivery, showing the number of impressions and optimization events as the ad is running.
Advertisers can also request this data be broken down along a number of different dimensions, including age, gender, and location (Designated Market Area~\cite{NeilsonDMARegions}, or DMA, region).
Notably, the interface and API {\em do not} provide a breakdown of ad delivery along racial lines; thus, analyzing delivery along racial lines necessitates development of a separate methodology that we describe in the next section.

\para{Anti-discrimination rules}
In response to issues of potential discrimination in online advertising reported by researchers and journalists~\cite{FacebookExcludeRace}, Facebook currently has several policies in place to avoid discrimination for certain types of ads.
Facebook also recently built tools to automatically detect ads offering housing, employment, and credit, and pledged to prevent the use of certain targeting categories with those ads.~\cite{FacebookAdsUpdate}.
Additionally, Facebook relies on advertisers to self-certify~\cite{FacebookDiscriminationSelfCertification} that they are not in violation of Facebook's advertising policy prohibitions against discriminatory practices~\cite{FacebookDiscriminatoryPracticesPolicy}.
More recently, in order to settle multiple lawsuits stemming from these reports, Facebook no longer allows age, gender, or ZIP code-based targeting for housing, employment or credit ads, and blocks other detailed targeting attributes that are ``describing or appearing to relate to protected classes''~\cite{FacebookHousingCreditEmployment, FacebookHousingLawsuitResponse, FacebookSpecialAdCategory}.

\subsection{Related work}

Next, we detail related work on algorithm auditing, transparency, and discriminatory ad targeting.

\para{Auditing algorithms for fairness}
Following the growing ubiquity of algorithms in daily life, a community formed around investigating their societal impacts~\cite{sandvig-2014-auditing}.
Typically, the algorithms under study are not available to outside auditors for direct examination; thus, most researchers treat them as ``black boxes'' and observe their reactions to different inputs.
Among most notable results, researchers have shown price discrimination in online retail sites~\cite{hannak-2014-ecommerce}, gender discrimination in job sites~\cite{hannak-2017-bias,chen-2018-chi}, stereotypical gender roles re-enforced by online translation services~\cite{bolukbasi-2016-man} and image search~\cite{kay-2015-unequal}, disparate performance on gender classification for Black women~\cite{buolamwini-2018-gender}, and political partisanships in search~\cite{kulshrestha-2017-quantifying,diakopoulos-2018-vote,robertson-2018-auditing}.
Although most of the work focused exclusively on the algorithms themselves, recently researchers began to point out that auditors should consider the entire socio-technical systems that include the users of those algorithms, an approach referred to as ``algorithm-in-the-loop''~\cite{green-2019-disparate,sapiezynski-2019-fairness}.
Furthermore, recent work has demonstrated that fairness is not necessarily composable, i.e., for several notions of fairness such as individual fairness~\cite{dwork-2012-fairness}, a collection of classifiers that are fair in isolation do not necessarily result in a fair outcome when they are used as part of a larger system~\cite{dwork-2018-fairness}.

\para{Advertising transparency}
In parallel to the developments in detecting and correcting unfairness, researchers have conducted studies and introduced tools with the aim of increasing transparency and explainability of algorithms and their outcomes.
For example, much attention has been dedicated to shedding light on the factors that influence the targeting of a particular ad on the web~\cite{lecuyer-2014-xray,lecuyer-2015-sunlight,parra-arnau-2017-myadchoices,EyeWnder} and on specific services~\cite{datta-2015-automated,wills-2012-ads}.

Focusing on Facebook, Andreou~\etal investigated the transparency initiative from Facebook that purportedly tells users why they see particular targeted ads~\cite{andreou-2018-explanations}.
They found that the provided explanations are incomplete and, at times, misleading.
Venkatadri~\etal introduced the tool called ``TREADS'' that attempts to close this gap by providing Facebook users with detailed descriptions of their inferred attributes using the ads themselves as a vehicle~\cite{venkatadri-2018-treads}.
Further, they investigated how data from third-party data brokers is used in Facebook's targeting features and---for the first time---revealed those third-party attributes to the users themselves using TREADS~\cite{venkatadri-2019-databrokers}.
Similar to other recent work~\cite{neumann-2018-profiling}, Venkatadri~\etal found that the data from third-party data brokers had varying accuracy~\cite{venkatadri-2019-databrokers}.

\para{Discrimination in advertising}
As described above, Facebook has some policies and tools in place to prevent discriminatory ad targeting. However, advertisers can still exclude users based on a variety of interests that are highly correlated with race by using custom audiences~\cite{speicher-2018-targeted}, or by using location~\cite{faizullabhoy-2018-fbattacks, korolova-2018-medium}.
Separately, Sweeney~\cite{sweeney-2013-discrimination} and Datta~\etal~\cite{datta-2015-automated} have studied discrimination in Google's advertising system.

The work just described deals with identifying possibilities for the advertisers to run discriminatory ads using the platform's features.
In contrast, other researchers, as well as and HUD's recent complaint, have suggested that discrimination may be introduced by the ad platform itself, rather than by a malicious advertiser~\cite{lambrecht-2018-algorithmic, datta-2015-automated, UpturnFacebookAmicusBrief, FacebookHUDLawsuitProPublica}.
For example, Lambrecht~\etal ran a series of ads for STEM education and found they were consistently delivered more to men than to women, even though there are more female users on Facebook, and they are known to be more likely to click on ads and generate conversions~\cite{lambrecht-2018-algorithmic}.
Datta~\etal explored ways that discrimination could arise in the targeting and delivery of job-related ads, and analyzed how different parties might be liable under existing law~\cite{datta-2018-discrimination}.
Our work explores these findings in depth, separating market effects from optimization effects and exploring the mechanisms by which ads are delivered in a skewed manner.

\section{Methodology}\label{sec:methodology}
We now describe our methodology for measuring the delivery of Facebook ads. 
At a high level, our goal is to run groups of ads where we vary a particular feature, with the goal of then measuring how changing that feature skews the set of users the Facebook platform delivers the ad to.
To do so, we need to carefully control which users are in our target audience.
We also need to develop a methodology to measure the ad delivery skew along racial lines, which, unlike gender, is not provided by Facebook's existing reporting tools.
We detail how we achieve that in the following sections.

\subsection{Audience selection}\label{sec:audiences}

When running ads, we often wish to control exactly which ad auctions we are participating in.  
For example, if we are running multiple instances of the same ad (\eg to establish statistical confidence), we do not want the instances to be competing against each other.
To this end, we use random PII-based custom audiences, where we randomly select U.S. Facebook users to be included in mutually-exclusive audiences.
By doing so, we can ensure that our ads are only competing against each other in the cases where we wish them to.
We also replicate some of the experiments while targeting all U.S. users to ensure that the effects do not only exist when custom audiences are targeted. 
As we show later in Section~\ref{sec:analysis}, we observe equivalent skews in these scenarios, which leads us to believe that preventing internal competition between our own ads is not crucial to measure the resulting skews.

\para{Generating custom audiences} 
We create each custom audience by randomly generating 20 lists of 1,000,000 distinct, valid North American phone numbers (\texttt{+1 XXX XXX XXXX}, using known-valid area codes). 
Facebook reported that they were able to match approximately 220,000 users on each of the 20 lists we uploaded.

Initially, we used these custom audiences directly to run ads, but while conducting the experiments we noticed that---even though we specifically target only North American phone numbers---many ads were delivered to users outside of North America.
This could be caused by users traveling abroad, users registering with fake phone numbers or with online phone number services, or for other reasons, whose investigation is outside the scope of this paper.
Therefore, for all the experiments where we target custom audiences, we additionally limit them to people located in the U.S.

\subsection{Data collection}
Once one of our ad campaigns is run, we use the Facebook Marketing API to obtain the delivery performance statistics of the ad every two minutes.
When we make this request, we ask Facebook to break down the ad delivery performance according to the attribute of study (age, gender, or location).
Facebook's response to each query features the following fields, among others, for each of the demographic attributes that we requested:
\begin{packed_itemize}
\item{\texttt{impressions}:} The number of times the ad was shown
\item{\texttt{reach}:} The number of unique users the ad was shown to
\item{\texttt{clicks}:} The number of clicks the ad has received
\item{\texttt{unique\_clicks}:} The number of unique users who clicked 
\end{packed_itemize}
Throughout the rest of the paper, we use the \texttt{reach} value when examining delivery; thus, when we report ``Fraction of men in the audience'' we calculate this as the \texttt{reach} of men divided by the sum of the \texttt{reach} of men and the \texttt{reach} of women (see Section~\ref{sec:binaries} for discussion on using binary values for gender). 

\begin{table*}
\centering
\ra{1.3}
\begin{tabular}{p{3.75cm}cccccccc}\toprule
\multirow{2}{*}{\bf DMA(s)~\cite{NeilsonDMARegions} } & \multicolumn{2}{c}{\bf \# Records ($A$)} & \phantom{ab} &  \multicolumn{2}{c}{\bf \# Records ($B$) } & \phantom{ab} & \multicolumn{2}{c}{\bf \# Records ($C$) }\\
\cmidrule{2-3} \cmidrule{5-6} \cmidrule{8-9}
 & {\bf White} & {\bf Black} && {\bf White} & {\bf Black} && {\bf White} & {\bf Black} \\
\midrule
\multirow{2}{*}{\parbox{3.75cm}{
Wilmington,\\
Raleigh--Durham\\
}} & \multirow{2}{*}{400,000} & \multirow{2}{*}{0} && \multirow{2}{*}{0} & \multirow{2}{*}{400,000} && \multirow{2}{*}{900,002} & \multirow{2}{*}{0}\\
& & & & & & & & \\
\multirow{2}{*}{\parbox{3.75cm}{
Greenville-Spartanburg,\\
Greenville-New Bern,\\
Charlotte, Greensboro
\\
}} & \multirow{2}{*}{0} & \multirow{2}{*}{400,000} && \multirow{2}{*}{400,000} & \multirow{2}{*}{0} && \multirow{2}{*}{0} & \multirow{2}{*}{892,097}\\
\phantom{ab} & & & & & & & & \\

\bottomrule
\end{tabular}
\label{stats:race}
\caption{Overview of the North Carolina custom audiences used to measure racial delivery.  We divide the most populated DMAs in the state into two sets, and create three audiences each with one race per DMA set.  Audiences $A$ and $B$ are disjoint from each other; audience $C$ contains the voters from $A$ with additional white voters from the first DMA set and Black voters from the second DMA set.   We then use the statistics Facebook reports about delivery by DMAs to infer delivery by race.}\label{stats:race}
\end{table*}

\subsection{Measuring racial ad delivery}\label{subsec:race-audiences}
The Facebook Marketing API allows advertisers to request a breakdown of ad delivery performance along a number of axes but it does not provide a breakdown based on race.
However, for the purposes of this work, we are able to measure the ad delivery breakdown along racial lines by using location (Designated Market Area, or DMA\footnote{Designated Market Areas~\cite{NeilsonDMARegions} are groups of U.S. counties that Neilson defines as ``market areas''; they were originally used to signify a region where users receive similar broadcast television and radio stations.  Facebook reports ad delivery by location using DMAs, so we use them here as well.}) as a proxy.

Similar to prior work~\cite{speicher-2018-targeted}, we obtain voter records from North Carolina; these are publicly available records that have the name, address, race, and often phone number of each registered voter in the state.
We partition the most populated North Carolina DMAs into two sets; for the exact DMAs, please see Table~\ref{stats:race}.
We ensure that each racial group (white and Black) from a set of DMAs has a matching number of records of the other group in the other set of DMAs. 
We sample three audiences ($A$, $B$, and $C$) that fit these constraints from the voter records and upload as separate Custom Audiences to Facebook.\footnote{Unfortunately, Facebook does not report the number of these users who match as we use multiple PII fields in the upload file~\cite{venkatadri-2018-targeting}.}
Audiences $A$ and $B$ are disjoint from each other; audience $C$ contains the voters from $A$ with additional white voters from the first DMA set and Black voters from the second DMA set.
We create audiences in this way to be able to test both ``flipped'' versions of the audiences ($A$ and $B$), as well as large audiences with as many users as possible ($C$); we created audience $B$ as large as possible (exhausting all voters who fit the necessary criteria), and sampled audience $A$ to match its size.
The details of the resulting audiences are shown in Table~\ref{stats:race}.

When we run ads where we want to examine the ad delivery along racial lines, we run the ads to one audience ($A$, $B$, or $C$).
We then request that Facebook's Marketing API deliver us results broken down by DMA.
Because we selected DMAs to be a proxy for race, we can use the results to infer which custom audience they were originally in, allowing us to determine the racial makeup of the audience who saw (and clicked on) the ad.
Note that in experiments that involve measuring racial skew all ads target the same target audience.
The limited number of registered voters does not allow us to create many large, disjoint custom audiences like we do in other experiments. 
However, as we show with ads targeting all U.S. users, internal competition does not appear to influence the results.

\subsection{Ad campaigns}
We use the Facebook Ad API described in Section~\ref{subsec:fb-impl} to create all ads for our experiments and to collect data on their delivery.
We carefully control for any time-of-day effects that might be present due to different user demographics using Facebook at different times of the day: for any given experiment, we run all ads at the same time to ensure that any such effects are experienced equally by all ads.
Unless otherwise noted, we used the following settings:
\begin{packed_itemize}
\item{\em Objective:} Consideration\textrightarrow Traffic\footnote{This target is defined as: Send more people to a destination on or off Facebook such as a website, app, or Messenger conversation.}
\item{\em Optimization Goal:} Link Clicks
\item{\em Traffic destination:} An external website (that depends on the ads run)
\item{\em Creative:} All of our ads had a single image and text relevant to the ad.
\item{\em Audience selection:} We use custom audiences for many of our ads, as described in Section~\ref{sec:audiences}, and further restrict them to adult (18+) users of all genders residing in the United States.  For other ads, we target all U.S. users age 18 or older.
\item{\em Budget:} We ran most ads with a budget of \$20 per day, and stopped them typically after six hours. 
\end{packed_itemize}

\subsection{Measuring and comparing audiences}
\label{sec:races}
We now describe the measurements we make during our experiments and how we compute their confidence intervals.

\para{Binary values of gender and race}\label{sec:binaries}
Facebook's marketing API reports ``female'', ``male'', and ``uncategorized'' as the possible values for gender.
Facebook's users self-report their gender, and the available values are ``female'', ``male'', and ``custom''.
The latter allows the user to manually type in their gender (with 60 predefined gender identities suggested through auto-complete functionality) and select the preferred pronoun from ``female - her'', ``male - him'', and ``neutral - them''.
Across our experiments, we observe that up to 1\% of the audiences are reported as ``uncategorized'' gender.
According to Facebook's documentation this represents the users who did not not list their gender.\footnote{https://www.facebook.com/business/help/151999381652364}
We do not know whether the ``uncategorized'' gender also features users with self-reported ``custom'' gender.
Thus, in this work we only consider the self-reported binary gender values of ``female'' and ``male''.

Further, when considering racial bias, we use the self-reported information from voter records.
The data we obtained has 7,560,885 individuals, with 93\% reporting their race as either Black or White.
Among those, less than 1\% report their ethnicity as ``Hispanic/Latino''.
Thus, in this work, we only target the individuals with self-reported race of White or Black.
However, when running our experiments measuring race (and targeting specific DMAs), we observe that a fraction ($\sim$10\%) of our ads are delivered to audiences outside of our predefined DMAs, thus making it impossible for us to infer their race.
This fraction remains fairly consistent across our experiments regardless of what we advertise, thus introducing the same amount of noise across our measurements.
This is not entirely unexpected, as we are targeting users directly, and those users may be traveling, may have moved, may have outdated information in the voter file, etc.

We do not claim that gender or race are binary, but choose to focus the analysis on users who self-reported their gender as ``female'' or ``male'' and race as ``Black'' or ``White''.
This way, we report the observable skew in delivery only along these axes. We recognize that delivery can be \emph{further} skewed with respect to gender of non-binary users and/or users of other races in a way that remains unreported in this work.

\para{Measuring statistical significance} 
Using the binary race and gender features, throughout this work, we describe the audiences by the fraction of male users and the fraction of white users.
We calculate the lower and upper limits of the 99\% confidence interval around this fraction using the method recommended by Agresti and Coull~\cite{agresti-1998-approximate}, defined in Equation~\ref{eq:single_99}:
\begin{equation}
\begin{aligned}
L.L. &= \frac{\hat{p}+\frac{z^2_{\alpha/{2}}}{2n}-z_{\alpha/2}\sqrt{\frac{\hat{p}(1-\hat{p})}{n}+\frac{{z^2_{\alpha/2}}}{4n^2}}}{1+{z^2_{\alpha/2}}/n},\\
U.L. &= \frac{\hat{p}+\frac{z^2_{\alpha/{2}}}{2n}+z_{\alpha/2}\sqrt{\frac{\hat{p}(1-\hat{p})}{n}+\frac{{z^2_{\alpha/2}}}{4n^2}}}{1+{z^2_{\alpha/2}}/n},
\end{aligned}
\label{eq:single_99}
\end{equation}
where $L.L.$ is the lower confidence limit, $U.L.$ is the upper confidence limit, $\hat{p}$ is the observed fraction of the audience with the attribute (here: male), $n$ is the size of the audience reached by the ad. 
To obtain the 99\% interval we set $z_{\alpha/{2}}=2.576$.
The advantage of using this calculation instead of the more frequently used normal approximation 
\begin{equation}
p\pm z_{\alpha/{2}} \sqrt{\frac{\hat{p}(1-\hat{p})}{n}}
\end{equation}
is that the resulting intervals fall in the $(0, 1)$ range.
Whenever the confidence intervals around these fractions for two audiences are non-overlapping, we can make a claim that the gender or racial makeups of two audiences are significantly different~\cite{cumming-2005-inference}.
However, the converse is not true: overlapping confidence intervals do not necessarily mean that the means are not different (see Figure 4 in~\cite{cumming-2005-inference} for explanation).
In this work we report all the results of our experiments but for easier interpretation emphasize those where the confidence intervals are non-overlapping.
We further confirm that the non-overlapping confidence intervals represent statistically significant differences, using the difference of proportion test as shown in  Equation~\ref{eq:zscore}:
\begin{equation}
Z = \frac{(\hat{p_1}-\hat{p_2})-0}{\sqrt{\hat{p}(1-\hat{p})(\frac{1}{n_1}+\frac{1}{n_2})}}
\label{eq:zscore}
\end{equation}
where $\hat{p_1}$ and $\hat{p_2}$ are the fractions of men (white users) in the two audiences that we compare, $n_1$ and $n_2$ are sizes of these audiences, and $\hat{p}$ is the fraction of men (white users) in the two delivery audiences combined.
All the results we refer to as statistically significant are significant in this test with a $Z$-score of at least $2.576$.
Finally, as we present in the Appendix, the comparisons presented are statistically significant also after the application of Bonferroni correction~\cite{cabin-2000-bonferroni} for multiple hypotheses testing.

Note that in experiments where we run multiple instances of an ad targeting disjoint custom audiences, the values of $\hat{p}$ and $n$ are calculated from the sums of reached audiences.

\section{Experiments}\label{sec:analysis}
In this section, we explore how an advertiser's choice of ad creative (headline, text, and image) and ad campaign settings (bidding strategy, targeted audience) can affect the demographics (gender and race) of the users to whom the ad is ultimately delivered. 

\subsection{Budget effects on ad delivery}\label{sec:market}
We begin by examining the impact that market effects can have on delivery, aiming to test the hypothesis put forth by Lambrecht \etal~\cite{lambrecht-2018-algorithmic}.
In particular, they observed that their ads were predominantly shown to men even though women had consistently higher click through rates (CTRs).
They then hypothesized that the higher CTRs led to women being more expensive to advertise to, meaning they were more likely to lose auctions for women when compared to auctions for men.

We test this hypothesis by running the same ad campaign with different budgets; our goal is to measure the effect that the daily budget alone has on the makeup of users who see the ads.
When running these experiments, we keep the ad creative and targeted audience constant, only changing the bidding strategy to give Facebook different daily limits (thus, any ad delivery differences can be attributed to the budget alone). 
We run an ad with daily budget limits of \$1, \$2, \$5, \$10, \$20, and \$50, and run multiple instances at each budget limit for statistical confidence.
Finally, we run the experiment twice, once targeting our random phone number custom audiences, and once targeting all users located in U.S.; we do so to verify that any effect we see is not a function of our particular target audience, and that it persists also when non-custom audiences are targeted.

\begin{figure}
	\centering
	\includegraphics[width=1\linewidth]{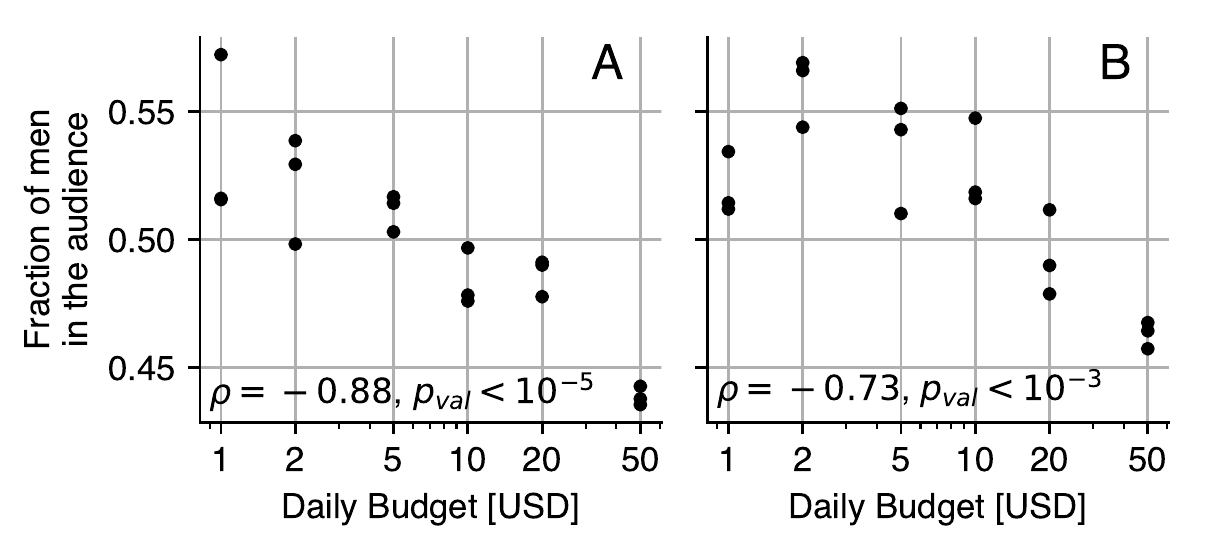}
	\caption{Gender distributions of the audience depend on the daily budget of an ad, with higher budgets leading to a higher fraction of women.  The left graph shows an experiment where we target all users located in the U.S.; the right graph shows an experiment where we target our random phone number custom audiences.}
	\label{fig:budget}
\end{figure}

Figure~\ref{fig:budget} presents the results, plotting the daily budget we specify versus the resulting fraction of men in the audience.
The left graph shows the results when we target all users located in the U.S., and the right graph shows the results when we target the random phone number custom audiences.
In both cases, we observe that changes in ad delivery due to differences in budget are indeed happening: the higher the daily budget, the smaller the fraction of men in the audience, with the Pearson's correlation of $\rho=-0.88$, $p_{val}<10^{-5}$ for all U.S. users and $\rho=-0.73$, $p_{val}<10^{-3}$ for the custom audiences.

The stronger effect we see when targeting all U.S. users may be due to the additional freedom that the ad delivery system has when choosing who to deliver to, as this is a significantly larger audience.

To eliminate the impact that market effects can have on delivery in our following experiments, we ensure that all runs of a given experiment use the same bidding strategy and budget limit. Typically we use a daily budget of \$20 per campaign.

\subsection{Ad creative effects on ad delivery}
Now we examine the effect that the ad creative (headline, text, and image) can have on ad delivery.
To do so, we create two stereotypical ads that we believe would appeal primarily to men and women, respectively: one ad focusing on \textit{bodybuilding} and another on \textit{cosmetics}.
The actual ads themselves are shown in Figure~\ref{fig:screenshot-bodybuilding}.
We run each of the ads at the same time and with the same bidding strategy and budget. 
For each variable we target different custom audiences, \ie the ``base'' level ads target one audience, ``text'' level ads target another, etc.
\textit{Note that we do not explicitly target either ad based on gender; the only targeting restrictions we stipulate are 18+ year old users in the U.S.}

We observe dramatic differences in ad delivery, even though the bidding strategy is the same for all ads, and each pair of ads target the same gender-agnostic audience.
In particular, the bodybuilding ad ended up being delivered to over 75\% men on average, while the cosmetics ad ended up being delivered to over 90\% women on average.
Again, this skewed delivery is despite the fact that we---the advertiser---did not specify difference in budget or target audience.

\para{Individual components' impact on ad delivery}
With the knowledge that the ad creative can skew delivery, we dig deeper to determine {\em which} of the components of the ad creative (headline, text, and image) have the greatest effect on ad delivery.
To do so, we stick with the bodybuilding and cosmetics ads, and ``turn off'' various features of the ad creative by replacing them with empty strings or blank images.
For example, the bodybuilding experiment listed as ``base'' includes an empty headline, empty ad text, and a blank white image; it does however link to the domain {\tt bodybuilding.com}.
Similarly, the cosmetics experiment listed as ``base'' includes no headline, text, or image, but does link to the domain {\tt elle.com}.
We then add back various parts of the ad creative, as shown in Figure~\ref{fig:screenshot-bodybuilding}.

\begin{figure}
	\centering
	\includegraphics[width=1\linewidth]{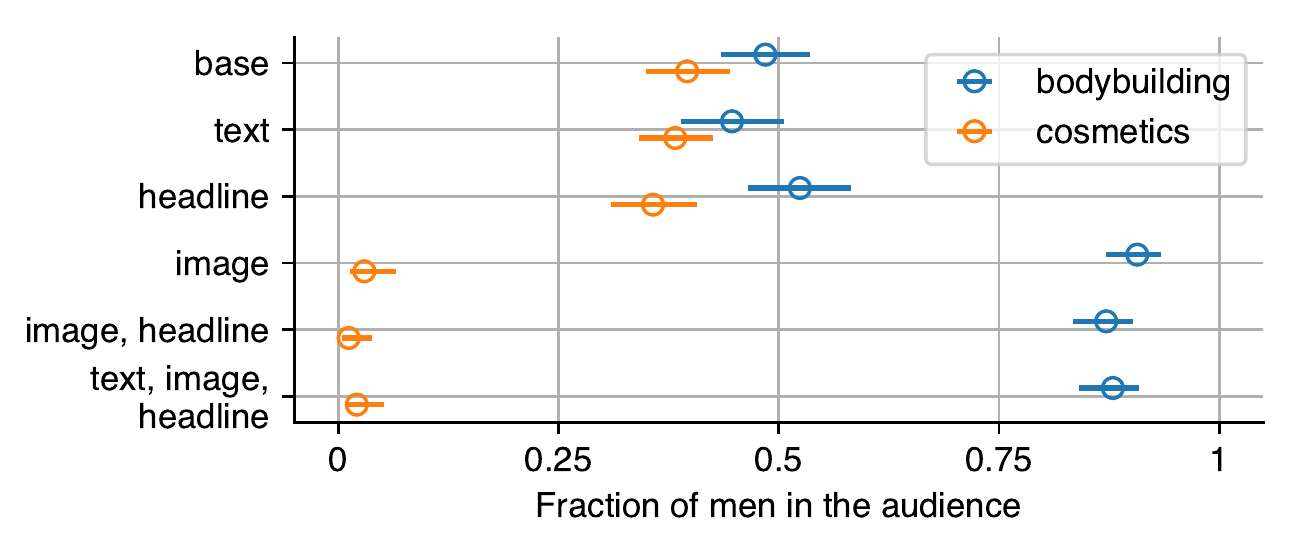}
	\caption{``Base'' ad contains a link to a page about either bodybuilding or cosmetics, a blank image, no text, or headline. 
	There is a small difference in the fraction of male users for the base ads, and adding the ``text'' only decreases it.
	Setting the ``headline'' sets the two ads apart but the audience of each is still not significantly different than that of the base version.
	Finally, setting the ad ``image'' causes drastic changes: the bodybuilding ad is shown to a 91\% male audience, the cosmetics ad is shown to a 5\% male audience, despite the same target audience.}
	\label{fig:classification-factors}
\end{figure}

The results of this experiment are presented in Figure~\ref{fig:classification-factors}.
Error bars in the figure correspond to 99\% confidence intervals as defined in Equation~\ref{eq:single_99}.
All results are shown relative to that experiment's ``base'' ad containing only the destination URL.
We make a number of observations.
{\em First}, we can observe an ad delivery difference due to the destination URL itself; the base bodybuilding ad delivers to 48\% men, while the base cosmetics ad delivers to 40\% men.
{\em Second}, as we add back the title and the headline, the ad delivery does not appreciably change from the baseline. 
However, once we introduce the image into the ad, the delivery changes dramatically, returning to the level of skewed delivery discussed above (over 75\% male for bodybuilding, and over 90\% female for cosmetics).
When we add the text and/or the headline back alongside the image, the skew of delivery does not change significantly compared to the presence of image only. 
Overall, our results demonstrate that the choice of ad image can have a dramatic effect on which users in the audience ultimately are shown the ad.

\para{Swapping images} 
To further explore how the choice of image impacts ad delivery, we continue using the bodybuilding and cosmetics ads, and test how ads with incongruent images and text are delivered.
Specifically, we swap the images between the two ads, running an ad with the bodybuilding headline, text, and destination link, but with the image from cosmetics (and vice versa).
We also run the original ads (with congruent images and text) for comparison.  

\begin{figure}
	\centering
	\includegraphics[width=1\linewidth]{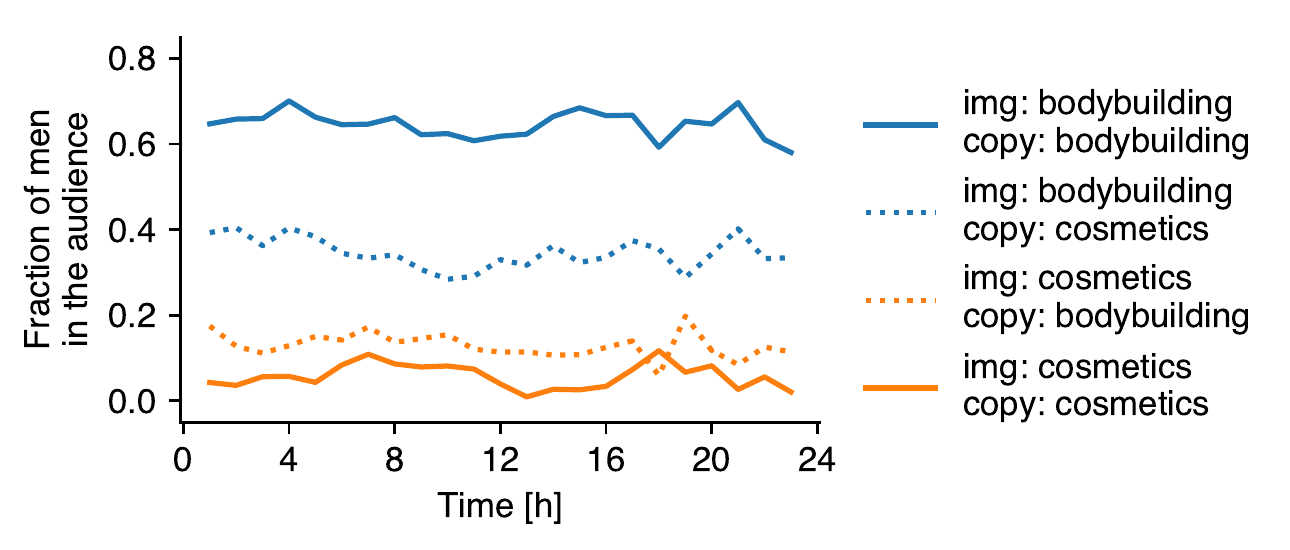}
	\caption{Ad delivery of original bodybuilding and cosmetics ads, as well as the same ads with incongruent images. Skew in delivery is observed from the beginning, and using incongruent images skews the delivery to a lesser degree.}
	\label{fig:image-switch}
\end{figure}

The results of this experiment are presented in Figure~\ref{fig:image-switch}, showing the skew in delivery of the ads over time.
The color of the lines indicates the image that is shown in the ad; 
solid lines represent the delivery of ads with images consistent with the description, while dotted lines show the delivery for ads where image was replaced.
We make a number of observations.
{\em First}, when using congruent ad text and image (solid lines), we observe the skew we observed before.
However, we can now see clearly that this delivery skew appears to exist from the very beginning of the ad delivery, \ie before users begin viewing and interacting with our ads. We will explore this further in the following section.
{\em Second}, we see that when we switch the images---resulting in incongruent ads (dotted lines)---the skew still exists but to a lesser degree.
Notably, we observe that the ad with an image of bodybuilding but cosmetics text delivers closest to 50:50 across genders, but the ad with the image of cosmetics but bodybuilding text does not.
The exact mechanism by which Facebook decides to use the ad text and images in influencing ad delivery is unknown, and we leave a full exploration to future work.

\para{Swapping images mid-experiment} 
Facebook allows advertisers to change their ad while it is running, for example, to update the image or text.
As a final point of analysis, we examine how changing the ad creative mid-experiment---after it has started running---affects ad delivery.
To do so, we begin the experiment with the original congruent bodybuilding and cosmetics ads; we let these run for over six hours.
We then swap the images on the running ads, thereby making the ads incongruent, and examine how ad delivery changes.

\begin{figure}
	\centering
	\includegraphics[width=1\linewidth]{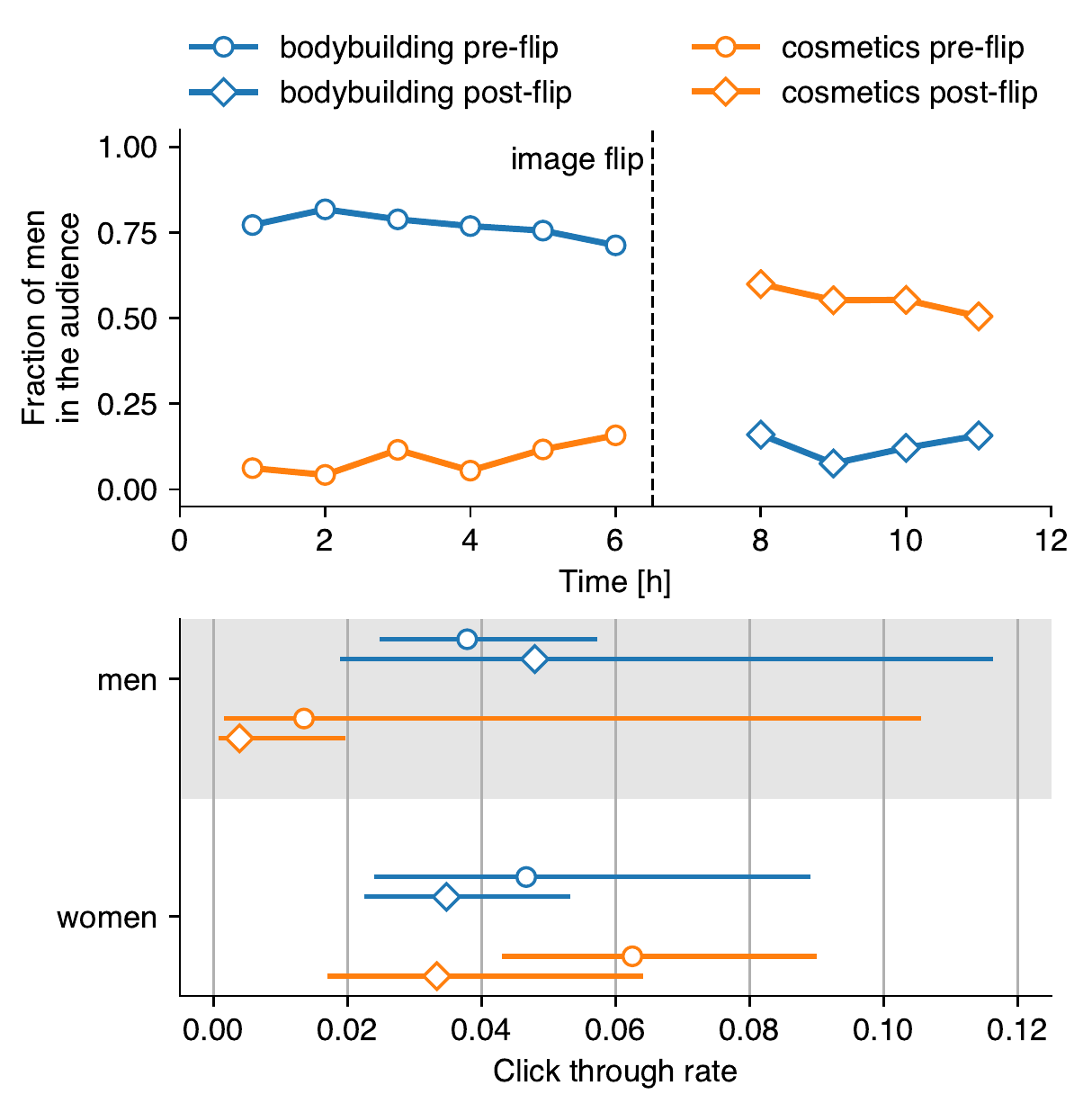}
	\caption{When we flip the image in the middle of the campaign, the ad is reclassified and shown to an updated audience. 
	Here, we start bodybuilding and cosmetics ads with corresponding descriptions and after 6 hours and 32 minutes we flip the images. 
	Within an hour of the change, the gender proportions are reversed, while there is no significant difference between the click through rates per gender pre and post flipping of the images.}
	\label{fig:image-flip}
\end{figure}

Figure~\ref{fig:image-flip} presents the results of this experiment.
In the top graph, we show the instantaneous ad delivery skew: as expected, the congruent ads start to deliver in a skewed manner as we have previously seen.
After the image swap at six hours, we notice a very rapid change in delivery with the ads almost completely flipping in ad delivery skew in a short period of time.
Interestingly, we do not observe a significant change in users' behavior to explain this swap: the bottom graph plots the click through rates (CTRs) for both ads by men and women over time. 
Thus, our results suggest that the change in ad delivery skew is unlikely to be due to the users' responses to the ads.

\subsection{Source of ad delivery skew}
We just observed that ads see a significant skew in ad delivery due to the contents of the ad, despite the bidding strategy and targeting parameters being held constant.
However, we observed that the ad delivery skew was present from the very beginning of ad delivery, and that swapping the image in the middle of a run resulted in a very rapid change in ad delivery that could not be explained by how frequently users click on our ads.
We now turn to explore the mechanism that may be leading to this ad delivery skew.

\para{Almost-transparent images}
We begin with the hypothesis that Facebook itself is automatically classifying the ad creative (including the image), and using the output of this classification to calculate a predicted relevance score to users.
In other words, we hypothesize that Facebook is running automatic text and image classification, rather than (say) relying on the ad's initial performance, which would explain (a) the delivery skew being present from the beginning of ad delivery, and (b) how the delivery changes rapidly despite no significant observable change in user behavior.
However, validating this hypothesis is tricky, as we are not privy to all of Facebook's ad performance data.

To test this hypothesis, we take an alternate approach.
We use the {\em alpha channel} that is present in many modern image formats; this is an additional channel that allows the image to encode the {\em transparency} of each pixel.
Thus, if we take an image and add an alpha channel with (say) 99\% opacity, all of the image data will still be present in the image, but any human who views the image would not be able to see it (as the image would show almost completely transparent).
However, if an automatic classifier exists, and if that classifier is not properly programmed to handle the alpha channel, it may continue to classify the image.

\newcommand{\img}[1]{\includegraphics[width=0.18\linewidth,height=0.18\linewidth]{#1}}
\newcommand{\imgno}[1]{\parbox[b][0.18\linewidth][c]{0.47cm}{~~~~#1}}
\begin{table}
\small
\begin{tabular}{c|cc|cc}
 & \multicolumn{2}{c|}{\bf Masculine} &  \multicolumn{2}{c}{\bf Feminine} \\
{\bf No.} & {\bf Visible} & {\bf Invisible} & {\bf Visible} & {\bf Invisible} \\
\hline
\imgno{1} & \img{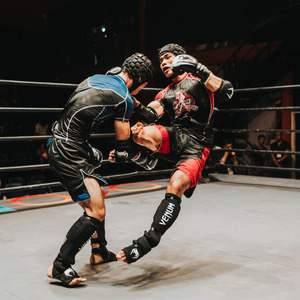} & \img{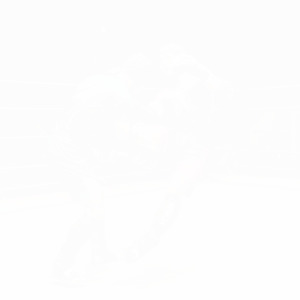} & \img{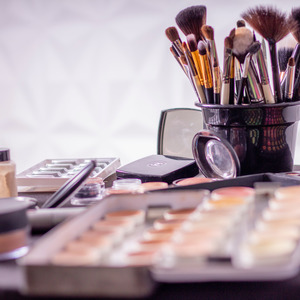} & \img{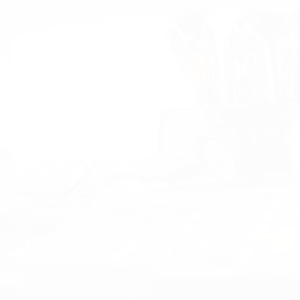} \\
\imgno{2} & \img{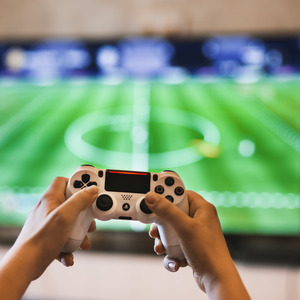} & \img{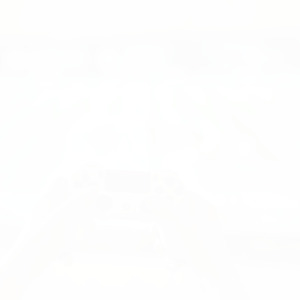} & \img{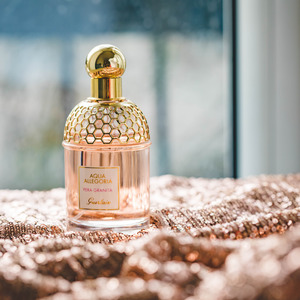} & \img{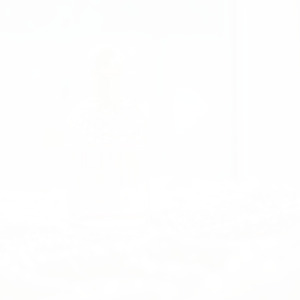} \\
\imgno{3} & \img{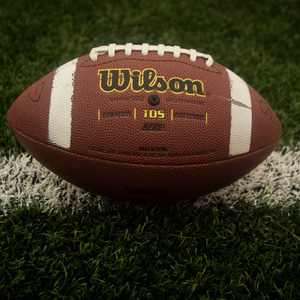} & \img{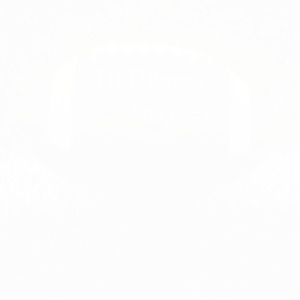} & \img{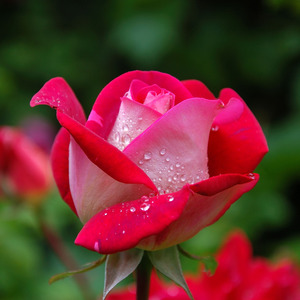} & \img{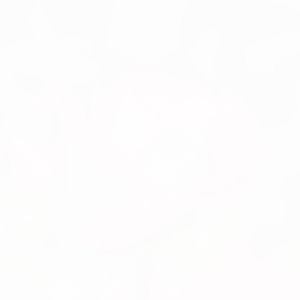} \\
\imgno{4} & \img{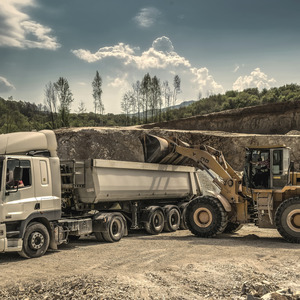} & \img{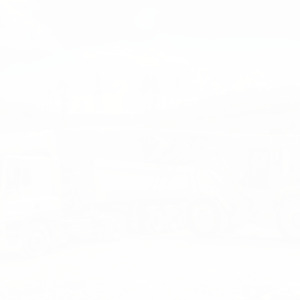} & \img{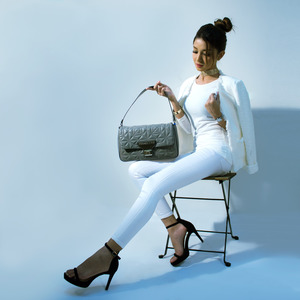} & \img{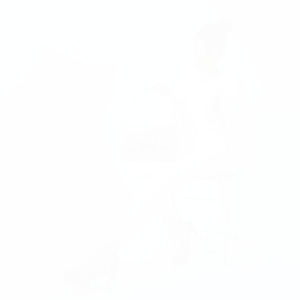} \\
\imgno{5} & \img{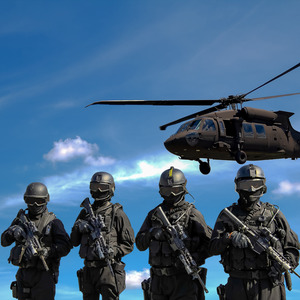} & \img{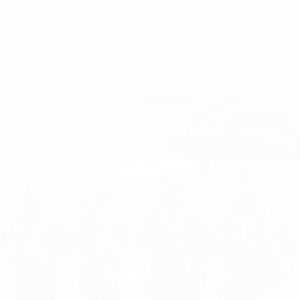} & \img{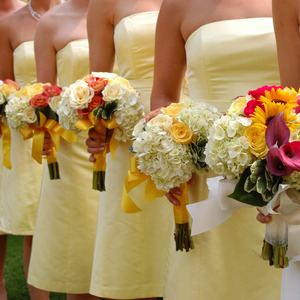} & \img{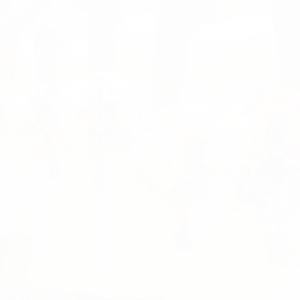} \\
\end{tabular}
\caption{Diagram of the images used in the transparency experiments.  Shown are the five stereotypical masculine and feminine images, along with the same images with a 98\% alpha channel, denoted as invisible.  The images with the alpha channel are almost invisible to humans, but are still delivered in a skewed manner.}\label{tab:images}
\end{table}

\para{Test images}
To test our hypothesis, we select five images that would stereotypically be of interest to men and five images that would stereotypically be of interest to women; these are shown in the second and fourth columns of Table~\ref{tab:images}.\footnote{All of these images were cropped from images posted to {\tt pexels.com}, which allow free non-commercial use.}$^,$\footnote{We cropped these images to the Facebook-recommended resolution of 1,080$\times$1,080 pixels to reduce the probability Facebook would resample the image.}
We convert them to PNG format add an alpha channel with 98\% opacity\footnote{We were unable to use 100\% transparency as we found that Facebook would run an image hash over the uploaded images and would detect different images with 100\% opacity to be the same (and would refuse to upload it again).  By using 98\% transparency, we ensure that the images were still almost invisible to humans but that Facebook would not detect they were the same image.} to each of these images; these are shown in the third and fifth columns of Table~\ref{tab:images}.
Because we cannot render a transparent image without a background, the versions in the paper are rendered on top of a white background.
As the reader can see, these images are not discernible to the human eye.

We first ran a series of tests to observe how Facebook's ad creation phase handled us uploading such transparent images.
If we used Reach as our ad objective, we found that Facebook ``flattened'' these images onto a white background in the ad preview.\footnote{Interestingly, we found that if we instead used Traffic as our ad objective, Facebook would both ``flatten'' these images onto a white background {\em and then normalize the contrast}.  This caused the ads to be visible to humans---simply with less detail that the original ads---thus defeating the experiment.  We are unsure of why Facebook did not choose to normalize images with the objective for Reach.}
By targeting ourselves with these Reach ads, we verified that when they were shown to users on the Facebook mobile app or in the desktop Facebook web feed, the images did indeed show up as white squares.
Thus, we can use this methodology to test whether there is an automatic image classifier present by examining whether running different transparent white ads results in different delivery.

\begin{figure}
	\centering
	\includegraphics[width=1\linewidth]{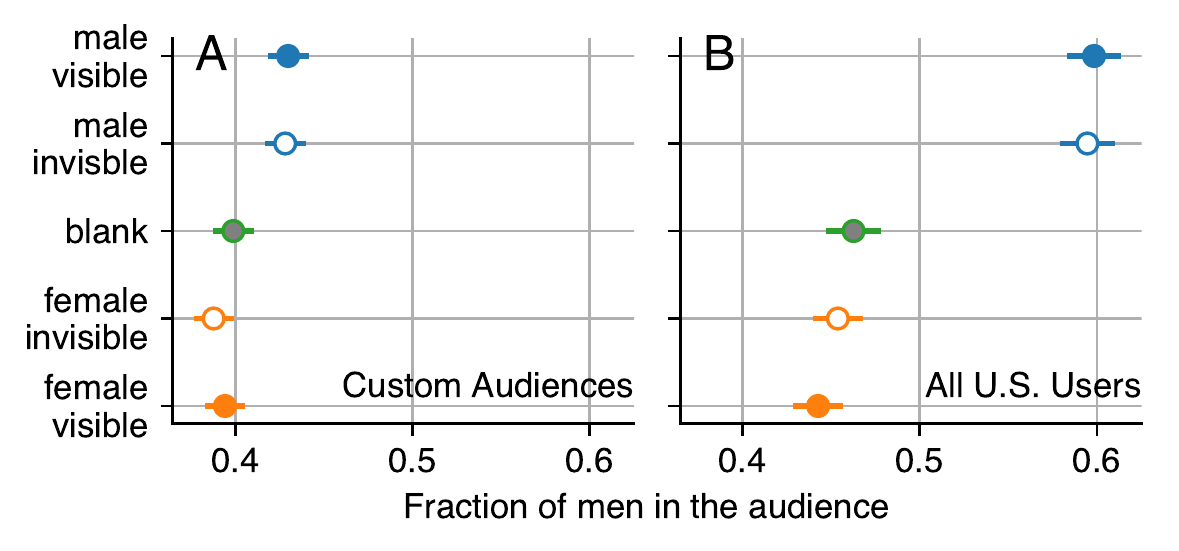}
	\caption{Fraction of reached men in the audiences for ads with the images from Table~\ref{tab:images}, targeting random phone number custom audience (A) and US audience (B). The solid markers are visible images, and the hollow markers are the same images with 98\% opacity. Also shown is the delivery to truly white images (``blank'').  We can observe that a difference in ad delivery exists, and that that difference is statistically significant between the masculine and feminine invisible images. This suggests that automated image classification is taking place.}
	\label{fig:image-transparency}
\end{figure}

\para{Results}
We run ads with all twenty of the images in Table~\ref{tab:images}, alongside ads with five truly blank white images for comparison.
For all 25 of these ads, we hold the ad headline, text, and destination link constant, run them all at the same time, and use the same bidding strategy and target custom audiences in a way that each user is potentially exposed to up to three ads (one masculine image, one feminine image, and one blank image).
We then record the differences in ad delivery of these 25 images along gender lines.
The results are presented in Figure~\ref{fig:image-transparency}A, with all five images in each of the five groups aggregated together.
We can observe that ad delivery is, in fact, skewed, with the ads with stereotypically masculine images delivering to over 43\% men and the ads with feminine images delivering to 39\% men in the experiment targeting custom audiences as well as 58\% and 44\% respectively in the experiment targeting all U.S. users.
Error bars in the plot correspond to the 99\% confidence interval calculated using Equation~\ref{eq:single_99}.

Interestingly, we also observe that the masculine invisible ads appear to be indistinguishable in the gender breakdown of their delivery from the masculine visible ads, and the feminine invisible ads appear to be indistinguishable in their delivery from the feminine visible ads.

As shown in Figure~\ref{fig:image-transparency}A, we verify that the fraction of men in the delivery of the male ads is significantly higher than in female-centered and neutral ads, as well as higher in neutral ads than in female-centered ads.
We also show that we cannot reject the null hypothesis that the fraction of men in the two versions of each ad (one visible, one invisible) are the same. 
Thus, we can conclude that the difference in ad delivery of our invisible male and female images is statistically significant, despite the fact that humans would not be able to perceive any differences in these ads.
This strongly suggests that our hypothesis is correct: that Facebook has an automated image classification mechanism in place that is used to steer different ads towards different subsets of the user population.\footnote{It is important to note we not know exactly how the classification works.  For example, the classifier may also be programmed to take in the ``flattened'' images that appear almost white, but there may sufficient data present in the images for the classification to work.  We leave a full exploration of how exactly the classifier is implemented to future work.}

To confirm this finding, we re-run the same experiment except that we change the target audience from our random phone number custom audiences (hundreds of thousands of users) to all U.S. users (over 320 million users).
Our theory is that if we give Facebook's algorithm a larger set of auctions to compete in, any effect of skewed delivery would be amplified as they may be able to find more users for whom the ad is highly ``relevant''.
In Figure~\ref{fig:image-transparency}B we observe that the ad delivery differences are, indeed, even greater: the male visible and invisible images deliver to approximately 60\% men, while the female visible and invisible images deliver to approximately 45\% men.
Moreover, the statistical significance of this experiment is even stronger, with a $Z$ value over 10 for the ad delivery difference between the male invisible and female invisible ads.

\subsection{Impact on real ads}
We have observed that differences in the ad headline, text, and image can lead to dramatic difference in ad delivery, despite the bidding strategy and target audience of the advertiser remaining the same.
However, all of our experiments thus far were on test ads where we typically changed only a single variable.
We now turn to examine the impact that ad delivery can have on realistic ads, where all properties of the ad creative can vary.

\begin{figure}
	\centering
	\includegraphics[width=1\linewidth]{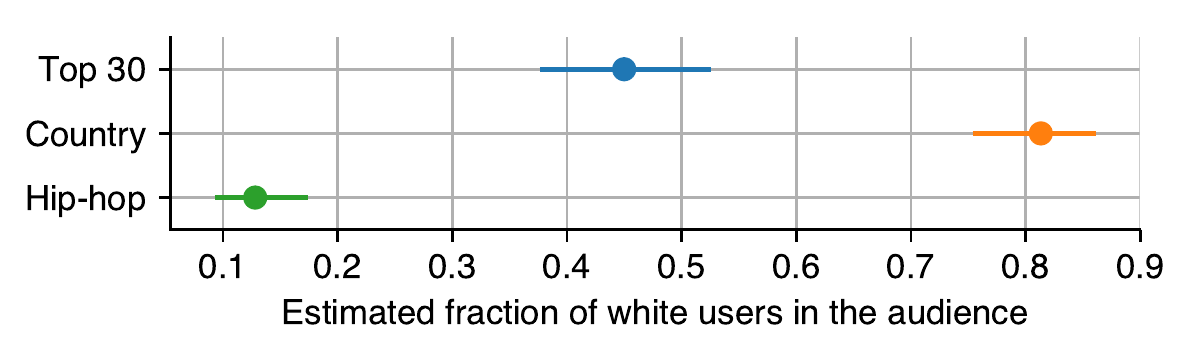}
	\caption{We run three campaigns about the best selling albums. \textit{Top 30} is neutral, targeting all. \textit{Country} implicitly targets white users, and \textit{Hip-hop} implicitly targets Black users. Facebook classification picks up on the implicit targeting and shows it to the audience we would expect.}
	\label{fig:image-race}
\end{figure}

\para{Entertainment ads}
We begin by constructing a series of benign entertainment ads that, while holding targeting parameters fixed (targeting custom audience $C$ from Table~\ref{stats:race}, are stereotypically of interest to different races.
Namely, we run three ads leading to lists of best albums in the previous year: general top 30 (neutral), top country music (stereotypically of interest mostly to white users), and top hip-hop albums (stereotypically of interest mostly to Black users).
We find that Facebook ad delivery follows the stereotypical distribution, despite all ads being targeted in the same manner and using the same bidding strategy.
Figure~\ref{fig:image-race} shows the fraction of white users in the audience in the three different ads, treating race as a binary (Black users constitute the remaining fraction).
Error bars represent 99\% confidence intervals calculated using Equation~\ref{eq:single_99}.

Neutral ads are seen by a relatively balanced, 45\% white audience, while the audiences receiving the country and hip-hop ads are 80\% and 13\% white, respectively.
Assuming significant population level differences of preferences, it can be argued that this experiment highlights the ``relevance'' measures embedded in ad delivery working as intended.
Next, we investigate cases where such differences may not be desired.

\begin{figure}
	\centering
	\includegraphics[width=1\linewidth]{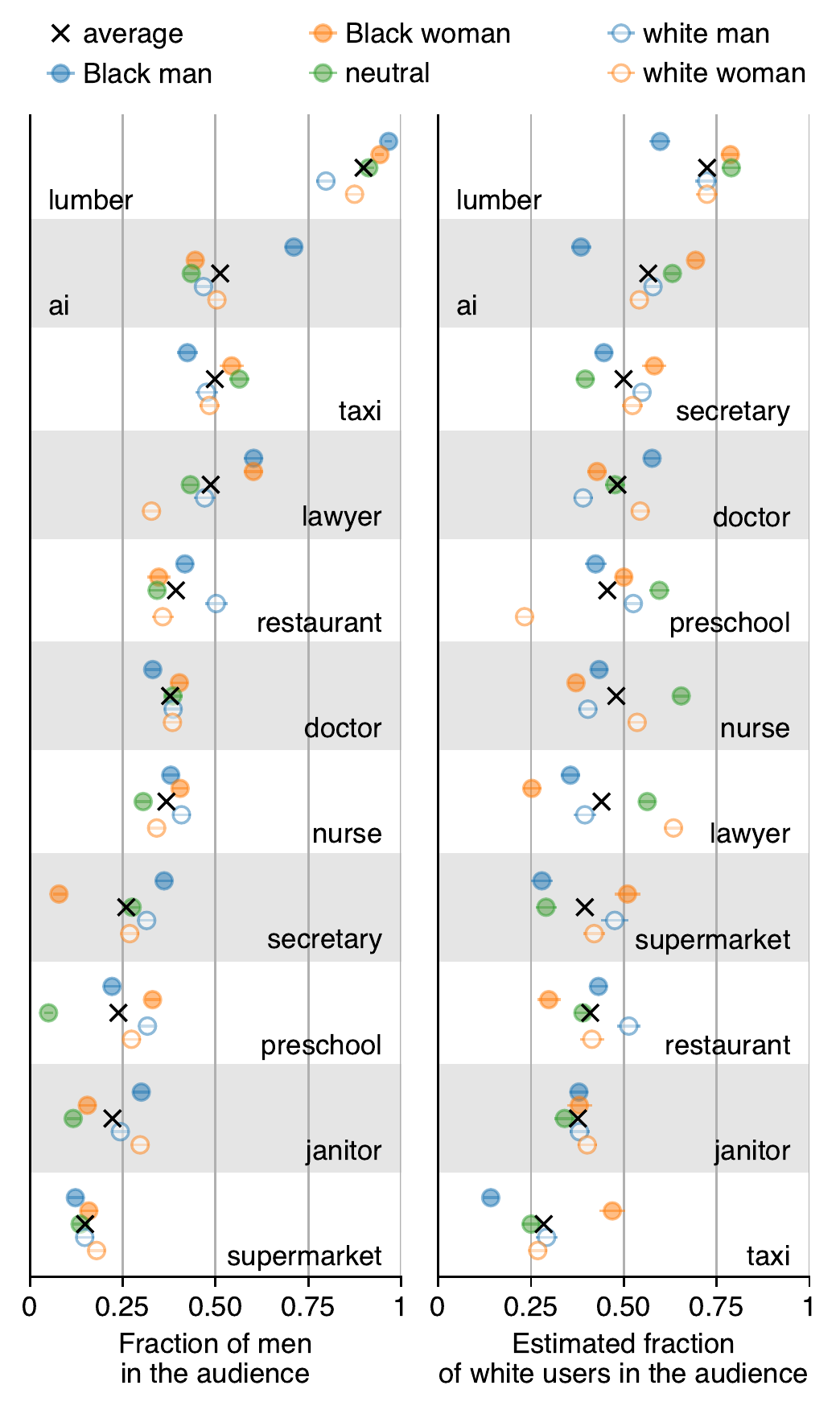}
	\caption{Results for employment ads, showing a breakdown of ad delivery by gender (left figure) and race (right figure) in the ultimate delivery audience.  The labels refer to the race/gender of the person in the ad image (if any).  The jobs themselves are ordered by the average fraction of men or white users in the audience.  Despite the same bidding strategy, the same target audience, and being run at the same time, we observe significant skew along on both racial and gender lines due to the content of the ad alone.}
	\label{fig:image-job-race}
\end{figure}

\para{Employment ads}
Next, we advertise eleven different generic job types: artificial intelligence developer, doctor, janitor, lawyer, lumberjack, nurse, preschool teacher, restaurant cashier, secretary, supermarket clerk, and taxi driver.
For each ad, we customize the text, headline, and image as a real employment ad would.
For example, we advertise for taxi drivers with the text ``Begin your career as a taxi driver or a chauffeur and get people to places on time.''
For each ad, we link users to the appropriate category of job listings on a real-world job site.

When selecting the ad image for each job type, we select five different stock photo images: one that has a white male, one that has a white female, one that has a black male, one that has a black female, and one that is appropriate for the job type but has no people in it.
We run each of these five independently to test a representative set of ads for each job type, looking to see how they are delivered along gender and racial lines (targeting custom audience $C$ from Table~\ref{stats:race}).
We run these ads for 24 hours, using the objective of Traffic, all targeting the same audience with the same bidding strategy.

The results of this experiment are presented in Figure~\ref{fig:image-job-race}, plotting the distribution of each of our ads along gender (left graph) and racial (right graph) lines.  
As before, the error bars represent the 99\% confidence interval calculated using Eq.~\ref{eq:single_99}.
We can immediately observe drastic differences in ad delivery across our ads along both racial and gender lines: our five ads for positions in the lumber industry deliver to over 90\% men and to over 70\% white users in aggregate, while our five ads for janitors deliver to over 65\% women and over 75\% black users in aggregate.
Recall that the only difference between these ads are the ad creative and destination link; we (the advertiser) used the same bidding strategy and target audience, and ran all ads at the same time.

Furthermore, we note that the skew in delivery cannot merely be explained by possibly different levels of competition from other advertisers for white and Black users or for male and female users. 
Although it is the case that each user may be targeted by a different number of advertisers with varying bid levels, by virtue of running all of our job ads at the same time, targeting the same users, with the same budget, we are ensuring that our ads are experiencing competition from other advertisers identically. 
In other words, our ad targeting asks that every user who is considered for our ``lumberjack'' job ad should also be considered for our taxi driver job ad, so these ads should be competing with each other and facing identical competition from other advertisers at auction time. 
If the content of the ad was not taken into account by the delivery optimization system, then the ads would be expected to have similar---though not necessarily even---breakdowns by race and gender at delivery. 
Our experiment demonstrates that this is not the case, and thus, aspects of ad delivery optimization, rather than merely advertiser competition, influence the skew in the delivery outcome. 

\para{Housing ads}
Finally, we create a suite of ads that advertise a variety of housing opportunities, as discrimination in online housing ads has recently been a source of concern~\cite{FacebookHUDLawsuit}.
We vary the type of property advertised (rental vs. purchase) and the implied cost (fixer-upper vs. luxury).
In each ad, the cost is implied through wording of the ad as well as the accompanying image.
Each ad points to a listing of houses for sale or rental apartments in North Carolina on a real-world housing site.
Simultaneously, we ran a baseline ad with generic (non-housing) text that simply links to \texttt{google.com}.
All of the ads ran for 12 hours, using the objective of Traffic, all targeting the same North Carolina audiences and using the same bidding strategy.
We construct the experiment such that each particular ad is run twice: once targeting audience $A$ and once targeting audience $B$ (see Table~\ref{stats:race})
This way we eliminate any potential geographical effects (for example, users in Wilmington could be interested in cheap houses to buy, and users in Charlotte could be interested in luxury rentals regardless of their ethnicity, but if we only used audience $C$ that effect could appear as racial skew).

\begin{figure}
	\centering
	\includegraphics[width=1\linewidth]{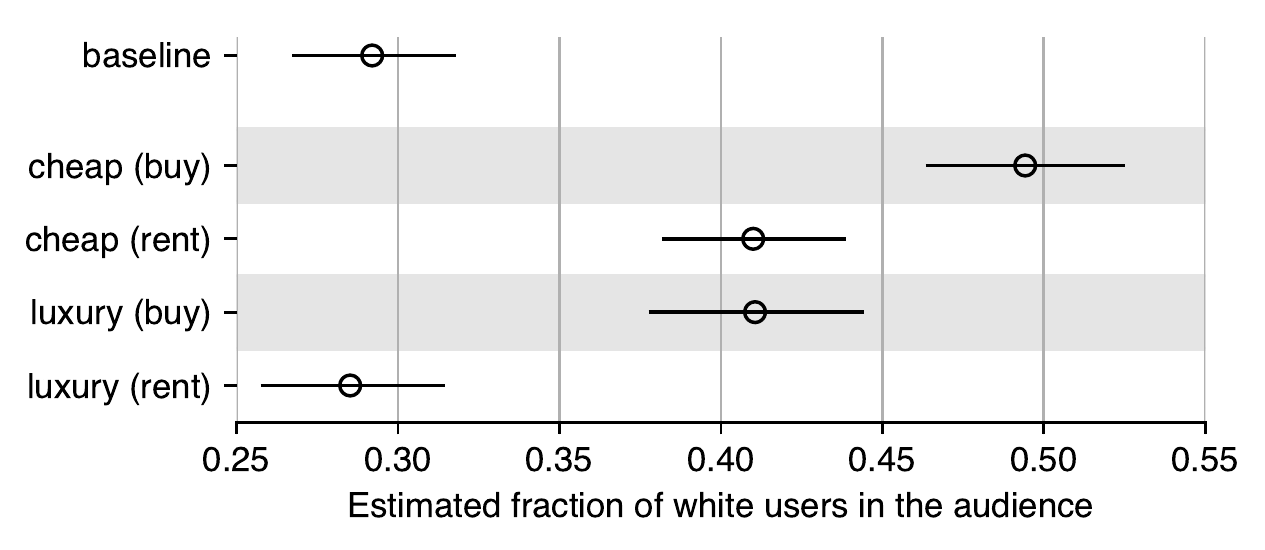}
	\caption{Results for housing ads, showing a breakdown in the ad delivery audience by race.  Despite being targeted in the same manner, using the same bidding strategy, and being run at the same time, we observe significant skew in the makeup of the audience to whom the ad is delivered (ranging from estimated 27\% white users for luxury rental ads to 49\% for cheap house purchase ads).  }
	\label{fig:image-housing-race}
\end{figure}

We present the results in Figure~\ref{fig:image-housing-race} (we found little skew for the housing ads along gender lines, and we omit those results).
We observe significant ad delivery skew along racial lines in the delivery of our ads, with certain ads delivering to an audience of over 72\% Black users (comparable to the baseline results) while others delivering to an audience of as little as 51\% Black users.

As with the employment ads, we cannot make claims about what particular properties of our ads lead to this skew, or about how housing ads in general are delivered.
However, given the significant skew we observe with our suite of ads, it indicates the further study is needed to understand how real-world housing ads are delivered.

\section{Concluding Discussion}\label{sec:discussion}
To date, the public debate and ad platform's comments about discrimination in digital advertising have focused heavily on the targeting features offered by advertising platforms, and the ways that advertisers can misuse those features~\cite{FacebookDoingMore}.

In this paper, we set out to investigate a different question: {\em to what degree and by what means may advertising platforms themselves play a role in creating discriminatory outcomes?}

Our study offers an improved understanding of the mechanisms behind and impact of ad delivery, a process distinct from ad creation and targeting.
While ad targeting is facilitated by an advertising platform---but nominally controlled by advertisers---ad delivery is conducted and controlled by the advertising platform itself.
We demonstrate that, during the ad delivery phase, advertising platforms can play an independent, central role in creating skewed, and potentially discriminatory, outcomes.
More concretely, we have:
\begin{itemize}
\item{} Replicated and affirmed prior research suggesting that market and pricing dynamics can create conditions that lead to differential outcomes, by showing that the lower the daily budget for an ad, the fewer women it is delivered to.
\item{} Shown that Facebook's ad delivery process can significantly alter the audience the ad is delivered to compared to the one intended by the advertiser based on the content of the ad itself. We used public voter record data to demonstrate that broadly and inclusively targeted ads can end up being differentially delivered to specific audience segments, even when we hold the budget and target audience constant.
\item{} Demonstrated that skewed ad delivery can start at the beginning of an ad's run.  We also showed that this process is likely automated on Facebook's side, and is not a reflection of the early feedback received from users in response to the ad, by using transparent images in ads that appear the same to humans but are distinguishable by automatic image classification tools, and showing they result in skewed delivery.
\item{} Confirmed that skewed delivery can take place on real-world ads for housing and employment opportunities by running a series of employment ads and housing ads with the same targeting parameters and bidding strategy.  Despite differing only in the ad creative and destination link, we observed skewed delivery along racial and gender lines.
\end{itemize}
We briefly discuss some limitations of our work and touch on the broader implications of our findings.

\para{Limitations}
It is important to note that while we have revealed certain aspects of how ad delivery is accomplished, and the effects it had on our experimental ad campaigns, we cannot make broad conclusions about how it impacts ads more generally.
For example, we observe that all of {\em our ads} for lumberjacks deliver to an audience of primarily white and male users, but that may not hold true of {\em all ads} for lumberjacks.
However, the significant ad delivery skew that we observe for our employment and housing ads strongly suggests that such skew is present for such ads run by real-world advertisers.

\para{Skew vs. discrimination}
Throughout this paper we refer to differences in the demographics of reached audience as ``skew'' in delivery.
We do not claim any observed skew {\em per se} is necessarily wrong or should be mitigated.
Without making value judgements on skew in general, we do emphasize the distinct case of ads for housing and employment.
In particular, the skew we observe in the delivery of ads for cosmetics or bodybuilding might be interpreted as reinforcing gender stereotypes but is unlikely to have legal implications.
On the other hand, the skew in delivery of employment and housing ads is potentially discriminatory in a legal sense.

Further, 
for the experiments involving ethnicity, we attempted to create equally sized audiences (50\% white and 50\% Black).
However, solely the fact that ads are not delivered to an evenly split audience does not indicate skew, as there might be differences in matching rates (what fraction of registered voters are active Facebook users) per ethnicity, or the groups could have different temporal usage patterns.
Only when we run two or more ads at the same time, targeting the same audience, and these ads are delivered with different proportions to white and Black users, do we claim we observe skew in delivery.

Our focus lies in understanding the extent to which the ad platform's delivery optimization, rather than merely its targeting tools and their use as implied by Facebook~\cite{FacebookDoingMore}, determine the outcomes of ad delivery, and on highlighting that demographic skews presently arise for certain legally protected categories in Facebook, even when the advertiser targets broadly and inclusively.

\para{Skew in traditional media}
Showing ads to individuals most likely to engage with them is one of the cornerstone promises of online ad platforms. 
While in traditional media---such as newspapers and television---advertisers can also place their ads strategically to reach particular kinds of readers or viewers, there are three significant differences with implications for fairness and discrimination when compared to advertising on Facebook. 

{\em First}, when advertising in traditional media, \emph{the advertiser} has the ability to purposefully advertise to a wide and diverse audience, and be assured that their ads will reach that audience. 
As we show in this work, this is not the case for advertising on Facebook. Even if the advertiser intends to reach a general and diverse audience, their ad can be steered to a narrow slice within that specified audience, that is skewed in unexpected or undesirable ways. 

{\em Second}, \emph{the individual's} agency to see ads targeted at groups they do not belong to is more severely limited in the hyper-targeted and delivery-optimized scenario of online ad platforms. 
In traditional media, an individual interested in seeing ads targeted to a different demographic than they belong to has to merely watch programming or read a newspaper that they are not usually a target demographic for.
On Facebook, finding out what ads one may be missing out on due to gender, race, or other characteristic inferred or predicted by Facebook is more challenging.
A particularly motivated user could change their self-reported gender but might find themselves discouraged from doing so because the account's gender information is always public. 
Other characteristics, such as race and net worth, are inferred by Facebook (or accessed via third-party companies~\cite{venkatadri-2019-databrokers}) rather than obtained through user's self-reported data, which makes them challenging to alter for the purposes of seeing ads.
Moreover, although users can remove some of their inferred interests using ad controls on Facebook, they have no ability to control {\em negative inferences} Facebook may be making about them. 
For example, Facebook may infer that a particular user is ``not interested in working at a lumber yard'', and therefore, not show this user ads for a lumberjack job even if the employer is trying to reach them.  As a result, Facebook would be excluding them from an opportunity in ways unbeknownst to the user and to the advertiser.

{\em Third}, \emph{public interest scrutiny} of the results of advertising is much more difficult in online delivery-optimized systems than in traditional media. 
Advertising in traditional media can be easily observed and analyzed by many members of society, from individuals to journalists, and targeting and delivery outside the expectation norms can be detected and called out by many. 
In the case of hyper-targeted online advertising whose delivery is controlled by the platform, such scrutiny is currently outside reach for most ads~\cite{MozillaInadequateAdArchive, KnightPublicInterest}.

\para{Policy implications}
Our findings underscore the need for policymakers and platforms to carefully consider the role of the optimizations run by the platforms themselves---and not just the targeting choices of advertisers---in seeking to prevent discrimination in digital advertising.

{\em First}, because discrimination can arise in ad delivery independently from ad targeting, limitations on ad targeting---such as those currently deployed by Facebook to limit the targeting features that can be used---will not address discrimination arising from ad delivery.
On the contrary, to the extent limiting ad targeting features prompts advertisers to rely on larger target audiences, the mechanisms of ad delivery will have an even greater practical impact on the ads that users see.

{\em Second}, regulators, lawmakers, and platforms themselves will need to more deeply consider whether and how longstanding civil rights laws apply to modern advertising platforms in light of ad delivery dynamics.
At a high level, U.S.~federal law prohibits discrimination in the marketing of housing, employment and credit opportunities.
A detailed consideration of these legal regimes is beyond the scope of this paper.
However, our findings show that ad platforms themselves can shape access to information about important life opportunities in ways that might present a challenge to equal opportunity goals.

{\em Third}, in the U.S., Section 230 of the Communications Decency Act (CDA) provides broad legal immunity for internet platforms acting as publishers of third-party content.
This immunity was a central issue in recently-settled litigation against Facebook, who argued its ad platform should be protected by CDA Section 230 in part because its advertisers are ``wholly responsible for deciding where, how, and when to publish their ads.''~\cite{FacebookOnuohaMotionToDismiss}
Our research shows that this claim is misleading, particularly in light of Facebook's role in determining the ad delivery outcomes.
Even absent unlawful behavior by advertisers, our research demonstrates that Facebook's own, independent actions during the delivery phase are crucial to determining how, when, and to whom ads are shown, and might produce unlawful outcomes.
These effects can be invisible to, and might even create liability for, Facebook's advertisers.

Thus, the effects we observed could introduce new liability for Facebook.
In determining whether Section 230 protections apply, courts consider whether an internet platform ``materially contributes'' to the alleged illegal conduct.
Courts have yet to squarely consider how the delivery mechanisms described in this paper might affect an ad platform's immunity under Section 230.

{\em Fourth}, our results emphasize the need for increased transparency into advertising platforms, particularly around ad delivery algorithms and statistics for real-world housing, credit, or employment ads. Facebook's existing ad transparency efforts are not yet sufficient to allow researchers to analyze the impact of ad delivery in the real world.

\para{Potential mitigations} 
Given the potential impact that discriminatory ad delivery can have on exposure to opportunities available to different populations, a natural question is how ad platforms such as Facebook may mitigate these effects. 
This is not straightforward, and is likely to require increased commitment and transparency from ad platforms as well as development of new algorithmic and machine learning techniques.
For instance, as we have demonstrated empirically in Section~\ref{sec:market} (and as~\cite{dwork-2018-fairness} have shown theoretically), skewed ad delivery can occur even if the ad platform refrains from refining the audience supplied by the advertisers according to the predicted relevance of the ad to individual users. This happens because different users are valued differently by advertisers, which, in a setting of limited user attention, leads to a tension between providing a useful service for users and advertisers, fair ad delivery, and the platform's own revenue goals.\footnote{A formal statement of this claim for the theoretical notions of individual fairness~\cite{dwork-2012-fairness} and its generalization, preference-informed fairness, can be found in \cite{kim-2019-fairness}.}

Thus, more advanced and nuanced approaches to addressing the potential issues of discrimination in digital advertising are necessary. Developing them is beyond the scope of this paper; however, we can imagine several different options, each with their own pros and cons. First, Facebook and similar platforms could disable optimization altogether for some ads, and deliver them to a random sample of users within an advertiser's target audience (with or without market considerations). Second, platforms could remove ads in protected categories from their normal ad flows altogether, and provide those listings in a separate kind of marketing product (\eg, a user-directed listing service like \texttt{craigslist.org}). Third, the platforms could allow the advertisers to enforce their own demographic outcomes so long as those desired outcomes don't themselves violate anti-discrimination laws. Finally, the platforms could seek to constrain their delivery optimization algorithms to satisfy chosen fairness criteria (some candidates for such criteria from the theoretical computer science community are individual fairness~\cite{dwork-2012-fairness} and preference-informed fairness~\cite{kim-2019-fairness}, but a broader discussion of appropriate criteria involving policymakers is needed).

Digital advertising increasingly influences how people are exposed to the world and its opportunities, and helps keep online services free of monetary cost. At the same time, its potential for negative impacts, through optimization due to ad delivery, is growing.
Lawmakers, regulators, and the ad platforms themselves need to address these issues head-on.

\section*{Acknowledgements}
We thank the anonymous reviewers for their helpful comments.
We also thank NaLette Brodnax and Christo Wilson for their invaluable feedback on the manuscript and Martin Goodson for pointing out erroneous confidence intervals.
The authors also acknowledge Hannah Masuga, a graduate fellow at Upturn, whose initial experiments during her fellowship inspired this research.
This work was funded in part by a grant from the Data Transparency Lab and NSF grant CNS-1616234.
This work was done in part while Aleksandra Korolova was visiting the Simons Institute for the Theory of Computing.

\section*{Errata}
\textbf{v2:} In the version of the paper published on April 3rd, 2019 we wrongly stated in Section~\ref{sec:races} that $\sim$40-50\% of ads were delivered outside of our predefined DMAs.
In version \textbf{v2} we corrected this figure to $\sim$10\%.
Further, in response to a request from Facebook, we changed the axis labels from ``Fraction of white users in the audience'' to ``Estimated fraction of white users in the audience''.

\noindent\textbf{v4:} We changed the method of calculating confidence intervals from normal approximation to the method described by Agresti and Coull~\cite{agresti-1998-approximate}. 
All confidence intervals presented in the figures throughout the paper use this method.
The change does not affect any of the conclusions. Notably, after the change the confidence intervals in Figure~\ref{fig:image-switch} no longer cross $0$.

\noindent\textbf{v5:} Includes the feedback and requested changes from anonymoys reviewers. The content is identical to this accepted at Computer-Supported Cooperative Work (CSCW) 2019.

\balance
\bibliographystyle{acm}

\bibliographystyle{acm --titlecase as-is --short month --no-field pages --sort key --sort author}

\appendix
\section*{Appendix}
\label{sec:appendix}
\begin{figure}
	\centering
	\includegraphics[width=0.99\linewidth]{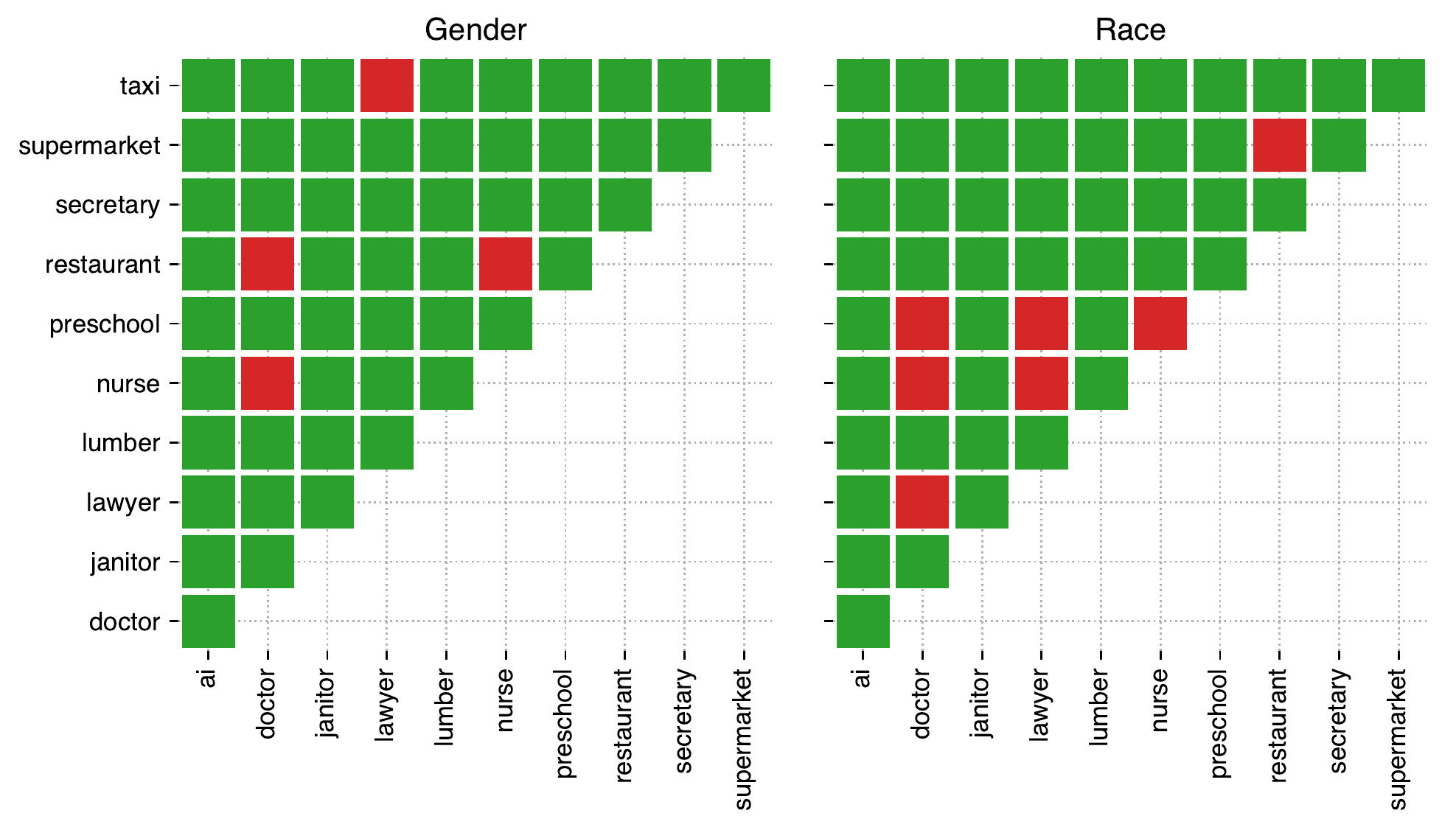}
	\caption{The demographic differences in ad delivery both in terms of gender and race are statistically significant after introducing Bonferroni correction with $N$ tests of 55. Green squares mark statistically significant differences, red squares indicate insignificant differences.}
	\label{fig:bonferroni}
\end{figure}

\para{Multiple hypotheses testing. } In the experiment described in the main paper we ran ads for 11 different job postings, each with five variations of the accompanying image. 
Here, we confirm that the apparent differences are not an effect of testing multiple hypotheses. We do so by aggregating the five variants for each ad and comparing the fraction of men and the estimated fraction of white users between each for pairs of jobs.
This results in 55 tests, so rather than using the $Z$ value corresponding to $p_{val}=0.01$, we use the Bonferroni correction~\cite{cabin-2000-bonferroni}, a statistical technique used to address the problem of making multiple comparisons. 
In Figure~\ref{fig:bonferroni} we show that the majority of comparisons remain statistically significant, each at the $Z$ value corresponding to corrected $p_{val}=\frac{0.01}{55}\approx{0.0002}$.

\newcommand{\etalchar}[1]{$^{#1}$}

\end{document}